\documentclass[twocolumn]{article}
\usepackage[utf8]{inputenc}
\usepackage[T1]{fontenc}
\usepackage{stfloats}
\usepackage{graphicx}
\usepackage{amsmath,amssymb,bm}
\usepackage[a4paper,margin=2cm]{geometry}

\usepackage{cuted}
\usepackage{capt-of}
\usepackage{natbib}

\begin{document}

\twocolumn[%
\begin{@twocolumnfalse}
\vspace*{-1em}

\begin{center}
{\LARGE Griffiths phase emerging from strong mutualistic disorder in high-dimensional interacting systems \par}
\vspace{0.8em}
{\large Tommaso Jack Leonardi$^{1}$, Amos Maritan$^{1}$, Sandro Azaele$^{1}$\par}
\vspace{0.4em}
$^{1}$Dipartimento di Fisica “G. Galilei”, Università di Padova, Italy
\end{center}

\vspace{0.5em}

\begin{abstract}
 The majority of analysis of interacting systems is done for weak and well-balanced interactions, when in fact topology and rare event factors often result in strong and sign-biased interactions when considering real systems. We analyse the impact of strong mutualistic interactions in a uniformly weighted Erd\H{o}s-Rényi network under generalised Lotka-Volterra dynamics. In difference to the typical case we show the interaction topology and system dynamics combine to produce power law abundance distributions in a critical region of the phase diagram, identifiable as a Griffiths phase. We find asymptotic expressions for the fixed point solution in this region, and establish the boundary of this region as when topology alone determines the abundance distribution. We show that the Griffiths phase persists for strong mutualistic interactions more generally, and survives when combined with weak all-to-all competition. 
\end{abstract}

\vspace{0.5em}
\end{@twocolumnfalse}%
]

Dynamical systems with many degrees of freedom are highly dependent on the degree of heterogeneity of the interactions between those variables. Empirically, it is common for systems with many components from ecology, economics and other fields to have scale-free or power law characteristics \cite{frankCommonPatternsNature2009,guimaraesStructureEcologicalNetworks2020}. Strong heterogeneity occurs when there is significant variation of magnitude of interactions between degrees of freedom, and when present in models of these different systems significantly alters behaviour.

In ecology, species abundance distributions commonly show non-Gaussian and power law behaviour, as do other macroecological patterns \cite{seguraMetabolicBasisFat2019,ser-giacomiUbiquitousAbundanceDistribution2018,azaeleStatisticalMechanicsEcological2016a,grilliMacroecologicalLawsDescribe2020}. However, this behaviour is not captured in the fixed point phase predictions from the typical Gaussian heterogeneity in complex system interactions used in the landmark works \cite{gallaDynamicallyEvolvedCommunity2018,buninEcologicalCommunitiesLotkaVolterra2017}, and even in this case, the interaction matrix corresponding to the surviving species is not well characterized by a Gaussian distribution \cite{baronBreakdownRandomMatrixUniversality2023a}. Real networks in fact often exhibit sparsity and structure \cite{newmanStructureRealworldNetworks2018}, and in ecosystems there are often selective and niche effects which can result in sparse interaction networks \cite{busielloExplorabilityOriginNetwork2017}, although other potential sources of the macroecological laws have been suggested \cite{descheemaekerHeavytailedAbundanceDistributions2021,gibbs2023can,suweisGeneralizedLotkaVolterraSystems2024a}.  

One approach to take more complex interaction networks into account gives effective statistical descriptions of high-dimensional time-varying systems, using dynamical mean-field theory and techniques in random matrix theory \cite{dedominicisDynamicsSubstituteReplicas1978a,gallaDynamicallyEvolvedCommunity2018}. This is classically done with all-to-all interactions but extended to general and sparse topologies \cite{aguirre-lopezHeterogeneousMeanfieldAnalysis2024a,metzDynamicalMeanFieldTheory2025,poleyInteractionNetworksPersistent2025,parkIncorporatingHeterogeneousInteractions2024d,allesina2012stability,khorunzhiyLifshitzTailsSpectra2006, cugliandoloMultifractalPhaseWeighted2024a,fisherCriticalBehaviorRandom1995,tonoloGeneralizedLotkaVolterraModel2025}. These methods allow for fine control of fluctuations by limiting the strength of the interactions, responding to questions concerning the persistent community and fixed point stability. 

There have been a variety of approaches to taking strong and sign-biased interactions into account. In the past it has been shown that strong disorder may be in general closely linked to the rare-region effects associated with Griffiths phases \cite{munozGriffithsPhasesComplex2010,morettiGriffithsPhasesStretching2013,brayGriffithsSingularitiesRandom1989}. Turning to sign, strong competition can lead to stationary or chaotic competitive-exclusion \cite{mallminChaoticTurnoverRare2024b,marcusLocalCollectiveTransitions2022,marcusLocalExtensiveFluctuations2024}, with topology informing when which subcommunities of species survive. In contrast, while mutualistic interactions can be seen as a source of instability due to the ease with which positive feedback causes unbounded growth \cite{rohrWillLargeComplex2025a}, they have been seen also to provide additional dynamic stability and diversity in consumer-resource models \cite{haleMutualismIncreasesDiversity2020}. 

In this Letter we first consider a minimal model for the role of strong mutualistic interactions in a sparse network, and obtain the exact phase diagram through generating function analysis. We find an emergent abundance power law phase for strong interactions, identifiable as a Griffiths phase \cite{griffithsNonanalyticBehaviorCritical1969}. We find the leading order series solution for the local field cumulant generating function in the case of Erd\H{o}s-Rényi (ER) networks, and establish the behaviour at the critical threshold. Finally we make arguments for criteria for the existence of the extended critical phase for general mutualistic interactions and show the robustness of this phase in presence of competitive interactions.

\textit{GLV Model}\textemdash
We consider the paradigmatic random generalised Lotka-Volterra model for $S$ species having corresponding abundances $x_{i}$

\begin{equation}\label{eq:glv}
    \dot{x}_{i}(t) = x_{i}(t)\left(1-x_{i}(t)+\sum_{j\neq i}^{S}\alpha_{ij}x_{j}(t)\right)
\end{equation}

We have set carrying capacities all equal to 1 for simplicity. Mutualistic interactions are defined to be when $\alpha_{ij}$ are non-negative, so interactions can only serve to increase the growth rate. This restriction means that if a fixed point is reached, it must be feasible, since at the fixed point for all species $x_{i}^{*} = 1+ \sum_{j\neq i}\alpha_{ij}x_{j}^{*}>0$, implying no extinctions. As such it will also be unique and globally stable \cite{stoneFeasibilityStabilityLarge2018a, gohGlobalStabilityManySpecies1977, akjouj2024complex}.

We illustratively take $\alpha_{ij}$ to be independent and identically distributed ER graph edges, with the following probability distribution.

\begin{equation}\label{eq:ER}
    \begin{split}
        P_{S}(\alpha_{ij}) = (1-\frac{c}{S})\delta(\alpha_{ij}) + \frac{c}{S}\delta(\alpha_{ij}-\beta)
    \end{split}
\end{equation}

with $c,\beta>0$. Either the interaction does not exist, or with probability $\frac{c}{S}$ it takes the value $\beta$. We choose the scaling of $c$ to be $1/S$ so that connectivity is independent of system size for realism in interaction sparsity, and this choice also ensures all cumulants of the interaction distribution scale identically as $1/S$. This provides us with a natural object to describe the full statistics of the interactions at the thermodynamic limit: their scaled cumulant generating function (CGF), $F_{\alpha}(z) \equiv \lim_{S\to\infty}S\ln\int d\alpha e^{-i\alpha z}P_{S}(\alpha)$. Accordingly we write the $k$th scaled cumulant as $\langle\alpha^{k}\rangle_{C}\equiv\frac{d^{k}F_{\alpha}(z)}{d(-iz)^{k}}\bigg\rvert_{z=0}$. The Gaussian limit is recovered when the scaled cumulants $\langle \alpha^{k}\rangle_{C}$ vanish for $k>2$, and here we are extending to a more general limit \footnote{This formalism is valid for $P_S(\alpha)$ when its scaled CGF corresponds to an infinitely divisible distribution; $P_{S}(\alpha)$ itself may not be. To be infinitely divisible the CGF ($F(z)$) of a sum of $S$ i.i.d. variables $X^{(S)}_{i}$ with CGF $F_{X}^{(S)}(z)$ is given by  $F(z) = S F_{X}^{(S)}(z)$}, and we note that there are many interaction distributions that can lead to the same scaled CGF. For our illustrative case the interaction distribution in Eq. \eqref{eq:ER} gives $F_{\alpha}(z)=c(\exp{(-i\beta z)}-1)$ and $\langle \alpha^{k}\rangle_{C}=c\beta^{k}$, the Poisson statistics.

Treating generating functions as analytic objects can allow for extraction of a surprising quantity of information about their corresponding series \cite{flajoletAnalyticCombinatorics2009}. While holomorphic characteristic functions and CGFs have a natural interpretation as exponential power series with moments or cumulants as coefficients, they can also provide asymptotic information about the probability distribution even when series expansions have zero radius of convergence through singularity analysis \cite{flajoletSingularityAnalysisGenerating1990}. 

The fixed point states of Eq. \eqref{eq:glv} are given by the effective single species equation $x^{*}(\eta)=(1+\eta)\Theta(1+\eta)$, for a local field or noise $\eta$ with statistics determined self-consistently. The distribution of $x^{*}$ at the fixed point is the species abundance distribution (SAD), $P_{x}(x)\equiv \langle \delta(x-x^{*}(\eta))\rangle_{\eta}$, which is defined only on the non-negative real axis (abundances cannot be negative), indicated by the presence of the Heaviside function $\Theta$. We can find the self-consistent statistics of $\eta$ in the fixed point regime by considering its CGF $F_{\eta}(z)$ (derivation in supplemental material \cite{supplementary})

\begin{equation}\label{eq:stat}
    \begin{split}
         F_{\eta}(z) &=  \langle F_{\alpha}(x z)\rangle_{x}
    \end{split}
\end{equation} 
where $\langle \cdot\rangle_x$ represents the average over the stationary solutions, $x^*$.  Eq. \eqref{eq:stat} matches the result from dynamical mean-field derivations in \cite{azaeleGeneralizedDynamicalMean2024,metzDynamicalMeanFieldTheory2025}, and is comparable in role to the self-consistency expressions in \cite{brayDiffusionSparselyConnected1988,rodgersDensityStatesSparse1988}. Due to Marcinkiewicz's theorem a CGF cannot be a polynomial of degree
greater than 2, that is, either all but the first 2 cumulants vanish or there are an
infinite number of nonvanishing cumulants \cite{lukacsCharacteristicFunctions1970}. In the former case naturally $F_{\eta}(z)$ retains the quadratic (Gaussian) form of $F_{\alpha}(z)$, while in the latter $F_{\eta}(z)$ may have a markedly different functional form. 

An immediate consequence of the eq. \eqref{eq:stat} is the condition for the unbounded growth in absence of extinction. The first derivative of \eqref{eq:stat} at $z=0$ gives $\langle\eta\rangle=\langle\alpha\rangle\langle x\rangle_{x}$, implying $\langle x \rangle_{x} = \frac{1}{1-\langle\alpha\rangle}$ and thus we get unbounded growth for $\langle\alpha\rangle\geq1$ \footnote{eq. \eqref{eq:stat} implies that $\langle x^*\rangle_x (1-\langle \alpha\rangle)= 1+(-1-\eta)\Theta(-1-\eta)$, which cannot be satisfied unless $\langle \alpha \rangle <1$, thus implying that the region $\langle \alpha\rangle \geq 1$ corresponds to the unbounded growth}. Beyond this there is no defined mean abundance which is an unphysical artefact of the model's simplicity: when mean interaction strength exceeds carrying capacity the system is unregulated.

Using the Poisson CGF for our interactions this gives us a closed functional equation for the local field CGF from which the SAD can be derived.

\begin{equation}\label{eq:functional}
    \begin{split}
        F_{\eta}(z,\beta) = c(\exp{(-i\beta z +F_{\eta}(\beta z,\beta))}-1)
    \end{split}
\end{equation}

where we show explicit dependence on both $z$ and $\beta$. For $\beta\ll 1$ the solution can be approximated as quadratic in $z$ and thus the Gaussian solution is recovered with mean $c\beta$ and variance $c\beta^2$. 
This functional equation admits an entire function as solution for a region in the phase diagram $c\beta<1, \ \beta<1$ (see the light grey region of Fig. \ref{fig:pd}). The first of these conditions is the boundary for unbounded growth as shown above.  For $\beta>1$ there are strong fluctuations in species abundances, originating in the interaction intensities and topology. In this regime a power series ansatz is insufficient for the solution of eq \eqref{eq:functional}, since coefficients of $z^{k}$ diverge at $c\beta^{k}=1$; the implications of which we will explore.

\begin{figure*}
\centering
\includegraphics[width=9cm]{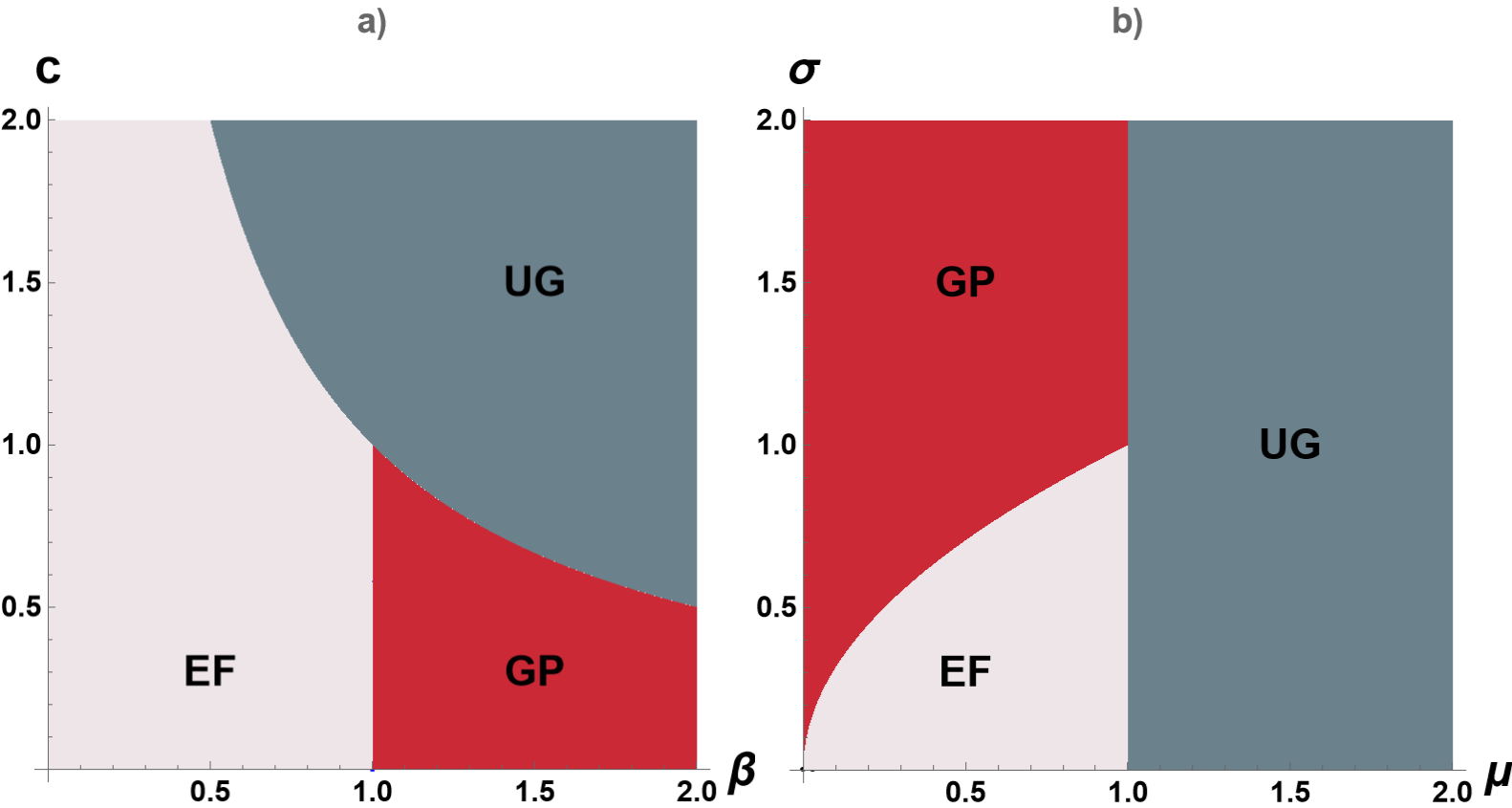}
\captionof{figure}{\label{fig:pd}Phase diagrams for the sparse GLV. Entire function (EF) and Griffiths phase (GP) have globally stable unique fixed points, with dividing line at $\beta=1$ as the onset of extended criticality. Unbounded growth (UG) when the DMFT mean of distribution is predicted to diverge ($c\beta=1$). a) in terms of finite connectivity $c$ and fixed interaction intensity $\beta$. b) in terms of overall interaction matrix element scaled mean, $\mu\equiv c\beta$ and scaled standard deviation, $\sigma \equiv \sqrt{c}\beta$, for comparison with the Gaussian case.} 
\end{figure*}

\textit{Solution for $c<1$ and $1<\beta<1/c$: the SAD is continuous and fat-tailed} \textemdash The full solution to eq. \eqref{eq:functional} has the form of a continued exponential, but for a tractable expression we can consider an approximation to eq. \eqref{eq:functional} valid for $c<1$, $\beta<\frac{1}{c}$

\begin{equation}\label{eq:functional approx}
    \begin{split}
        F_{\eta}(z) = ce^{-i\beta z}(1+F_{\eta}(\beta z))-c
    \end{split}
\end{equation}

which has series solution, 

\begin{equation}\label{eq:series sol}
    F_{\eta}(z) = -c + (1-c)\sum_{n=1}^{\infty}c^n\exp{(-iz \sum_{j=1}^{n}\beta^{j})}
\end{equation}

This is also the first contribution to the full solution for Eq.\eqref{eq:functional} with errors in each term of the sum in $n$ of order $c^{n+1}$ \cite{supplementary}.

For $c\beta\geq1$, where the first derivative is undefined and so in the unbounded growth phase, this expression recalls the Weierstrass function, everywhere continuous but not differentiable (see figure \ref{fig:fractal}). Via a Fourier transform it is possible to extract directly from this function the solution's spectral dimension \cite{mandelbrotFractalsFormChance1977}, which gives the exponent of the SAD to be $1-\frac{\ln{c}}{\ln{\beta}}$ \cite{supplementary}. 

To corroborate this exponent with a method whose validity extends beyond the series approximation for the ER case we retrieve it directly from the functional Eq.\ref{eq:functional}. For more details see \cite{supplementary}

We illustrate this method by expanding the functional equation around $z=0$ to find the local behaviour of $F_\eta(z)$ in the subregion $1/c^2 <\beta<1/c$, still with $0<c<1$. In this case $F_{\eta}(z)=-i \langle \eta\rangle z+a_{p}z^{p}+\mathcal{O}(z^{2})$. The exponent $p$ is given by condition $c\beta^{p}=1$ arising from the substitution of this $F_{\eta}(z)$ expansion into Eq. \eqref{eq:functional}: as a result $p= -\frac{\ln{c}}{\ln{\beta}}$. The mean local field is $\langle\eta\rangle = \frac{c\beta}{1-c\beta}$, found self-consistently, and due to the non-analyticity all higher cumulants are undefined. The coefficient $a_{p}$ cannot be locally determined \cite{jackiwHowSuperrenormalizableInteractions1981}, but all higher order contributions can be found from further local expansion and this coefficient.

The term $z^{p}$ is the largest order non-analyticity around $z=0$, and thus can be used to find the leading SAD behaviour as $x\to \infty$.

\begin{equation}\label{eq: disc/cont sad}
    \begin{split}
        P_{x}(x)  \sim x^{-(1-\frac{\ln{c}}{\ln{\beta}})}
    \end{split}
\end{equation}

This reasoning extends to the complete region $1<\beta<1/c$, where the exponent remains $p=-\frac{\ln{c}}{\ln{\beta}}$. The local expansion of $F_{\eta}(z)$ adjusts as $p$ varies to include analytic terms for all integer powers of $z$ up to $\lfloor p \rfloor$, each corresponding to finite cumulants of $\eta$. Higher order terms  after the non-analytic term $z^{p}$ have no direct interpretation. If $p$ itself is integer, $z^{p}\ln{z}$ is the largest order non-analyticity present in $F_{\eta}(z)$, giving a smooth change of exponent in the tail of $P_{x}(x)$.

These weak non-analyticities are examples of Griffiths singularities, characterised by the aggregation of exponentially rare events which in this case are the conjunctions of strong interactions \cite{vojtaRareRegionEffects2006,brayNatureGriffithsPhase1987,odorLocalizationTransitionLifschitz2014}. Griffiths found that these singularities exist for an extended region of temperature (here $1/\beta$), unlike the typical ferromagnet phase transition at a single critical point \cite{griffithsNonanalyticBehaviorCritical1969}. This extended region of critical behaviour, where the SAD tail is a pure power law with continuously varying exponent $p$, is more robust compared to typical critical point behaviour as it does not depend on fine-tuning of system parameters \cite{munozColloquiumCriticalityDynamical2018,grinsteinGenericScaleInvariance1991}. Thermodynamic Griffiths singularities in classical systems are very unlikely to be experimentally observed \cite{imryGriffithsSingularityFinite1977}, which here could mean a significantly high number of species and sample populations would be needed to identify the presence of this phenomenon.

\begin{figure*}
\centering
\includegraphics[width=16cm]{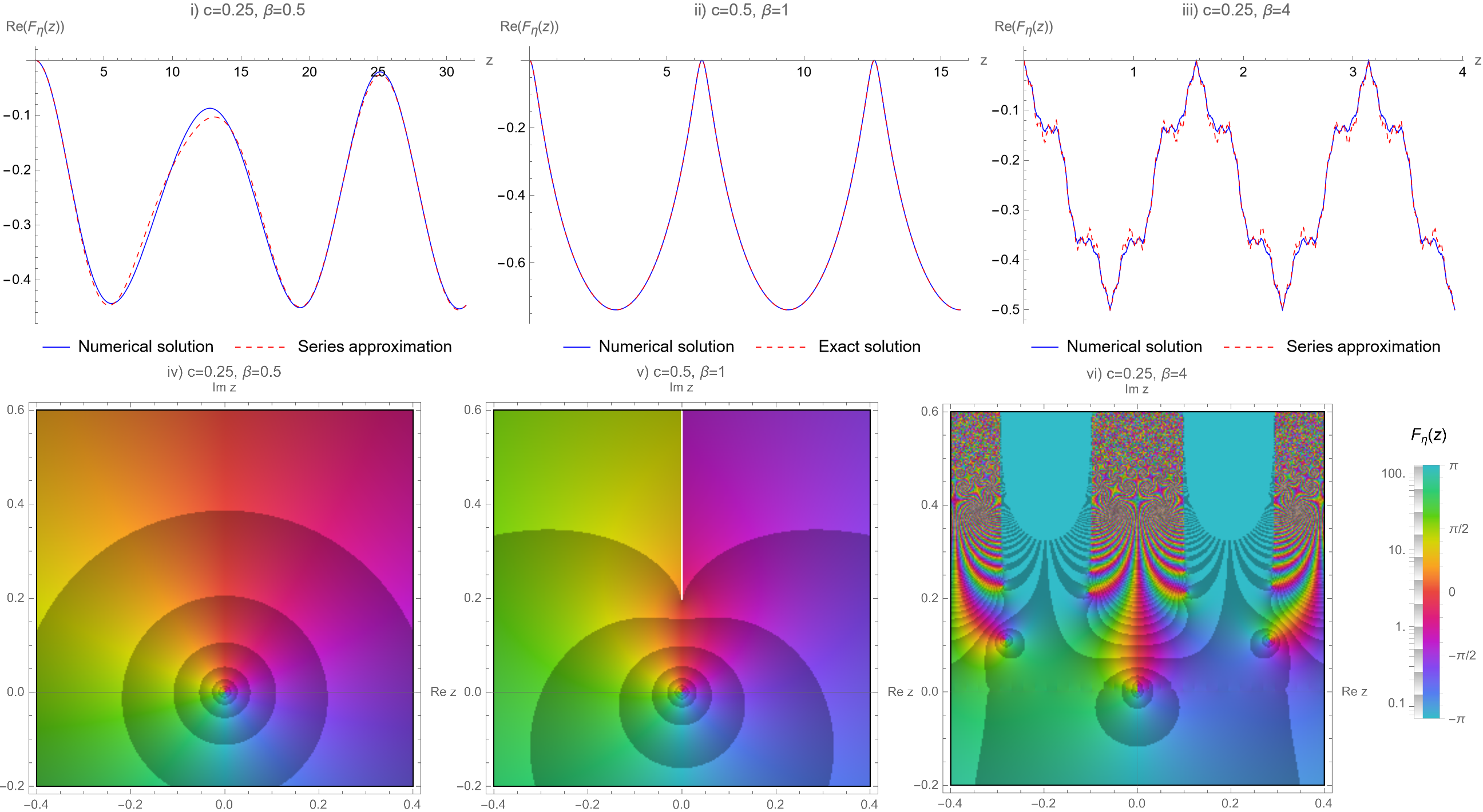}
\captionof{figure}{\label{fig:fractal}Solutions to Eq. \eqref{eq:functional} for the CGF of the local field at the fixed point, $F_\eta(z)$, to which the SAD is directly related. For varying interaction connectivity $c$ and intensity $\beta$, i)-iii) show $\text{Re}(F_\eta(z))$ against $z \in \mathbb{R}$, comparing numerical solutions with analytical results, while iv)-vi) show numerical computation of $|F_{\eta}(z)|$ (shading) and $\text{arg}(F_{\eta}(z))$ (hue) on $z\in \mathbb{C}$ around the origin.  i): $\beta<1$ case with $F_\eta(z)$ shown fully in iv); the solution is an entire function which is well approximated near the origin by a single exponential. ii): $\beta=1$ exact solution Eq. \eqref{eq:one} plotted and $F_{\eta}(z)$ shown fully in v), with behaviour impacted by the branch point at $i(c-1-\ln{c})$, a property of the ER network. As $c$ goes from $0$ to $1$ the branch point, i.e. the topology, increasingly dominates, giving pure semicircular waves on $z \in \mathbb{R}$ at $c=1$. iii): $\beta>1$ shows recursive fractal structure, on the real line reminiscent of the Weierstrass function (see Eq. \eqref{eq:series sol}). Formally the CGF cannot be defined on the upper half plane, so vi) shows a convergent approximation and not full detail.}
\end{figure*}

\textit{Solution for $c\leq 1$ and $\beta=1$: the SAD is discrete and is an exponentially truncated power law} \textemdash We note particular behaviour at exactly $\beta=1$ originating in the topology of the interactions. In this case we see the solution to the functional equation is

\begin{equation}\label{eq:one}
    \begin{split}
        F_{\eta}(z,\beta=1) = -c-W(-ce^{-c-iz})
    \end{split}
\end{equation}

Where $W(z)$ denotes the Lambert W function for which $W(z) e^{W(z)}=z$. We select the principal branch to satisfy normalization ($F_\eta(z=0)=0$), and imply this choice in future uses of the function.

This CGF corresponds to the following lattice probability distribution, where abundances can only take discrete integer values $x\geq1$ \footnote{If the characteristic function of a variable $\eta$, $\phi(z) \equiv e^{F_\eta(z)}$, satisfies $|\phi(z_0)|=1$ for some real $z_0$ and $|\phi(z)|< 1$ for $0<z<z_0$, then the density distribution $P_\eta(\eta)$ is concentrated on a lattice of values of $\eta$ where the lattice spacing is given by $2\pi/z_0$ \cite{breimanProbability1968}.}.

\begin{equation}\label{eq:borel}
    \begin{split}
        P_{x}(x)&= \frac{e^{-cx}(cx)^{x-1}}{x!} \ \ \ \ x \in \mathbb{N}_{+}
    \end{split}
\end{equation}

This is the Borel distribution, which is precisely the distribution of tree sizes in the sparse regime of the ER graph \cite{alonProbabilisticMethod2016}. This means there is direct correspondence between the topology of the interaction matrix and the resultant fixed point abundance distribution in this case where interaction strength $\beta$ is equal to carrying capacity $1$ \cite{munozGriffithsPhasesComplex2010}.  

Looking at large abundances ($x\to\infty$) where $x$ can be treated as a continuous variable, and using Stirling's approximation, will lead to the SAD tails
\begin{equation}\label{eq:one sad}
    \begin{split}
        P_{x}(x) \sim c^{-1}x^{-3/2}e^{-x(c-1-\ln{c})}
    \end{split}
\end{equation}

This can be recovered directly from the square-root singularity of eq. \eqref{eq:one}, as done in \cite{brayDiffusionSparselyConnected1988}. 

We see there is a universal exponent of $3/2$ for the regime $c<1, \ \beta=1$, with an exponential cutoff that vanishes at $c=1$. Thus at this latter point the topological signal in the abundances is most strong, and can be detected even for $\beta$ near $1$.

\textit{Extensions: General mutualistic intensity distributions}\textemdash A natural extension which remains amenable to analysis is to consider when the interaction intensities can take multiple values. For simplicity we begin with two $\beta_{1},\beta_{2}>0$: $P_{S}(\alpha)=\left(1-\frac{c}{S}\right)\delta(\alpha)+\frac{c}{S}\left(q_{1}\delta(\alpha-\beta_{1})+q_{2}\delta(\alpha-\beta_{2})\right)$ with weights $q_{1}+q_{2}=1$ for normalisation. 

The functional equation for this system is simply a sum weighted by $q_{i}$ of the single intensity version. From this we found both the first order series approximation and the expansion around $z=0$. The same non-analyticities are present, with exponent $p$ now given by $c (q_{1}\beta_{1}^{p}+q_{2}\beta_{2}^{p}) = 1 $. The finite cumulants of $\eta$ and higher order coefficients have more complicated expressions but can be found with the same methods as the single intensity case.

This analysis remains valid for $n$ different intensities $\beta_{n}$ with normalised weights $q_{n}$ and correspondingly in the continuum. This equates to defining the interaction distribution as a compound Poisson \cite{azaeleGeneralizedDynamicalMean2024}.  Here an intensity distribution $P_{\beta}(\beta)$ is defined which gives exponents of the SAD through inverting the relation $c \langle \beta^{k}\rangle =1$ where $\langle \beta^{k}\rangle$ are the moments of $P_{\beta}$. We give the series approximation and more details in the \cite{supplementary}.

So we see that the simplest case of the uniformly weighted ER interaction network gives an informative perspective for this class of systems allowing for a distribution of weights. The same non-analyticities are found in the CGF, with the exponent in the Griffiths phase now depending on the statistics of the intensities.

For generic mutualistic interactions ($\left(\int_{0}^{\infty}d\alpha P_{S}(\alpha)=1\right)$ unconstrained to ER topology, we find that the condition of $F_{\eta}(z)$ becoming non-analytic at $\langle \alpha^{k}\rangle_{C}=1$ for integer $k$ remains, indicating the Griffiths phase. As such we suggest that if $\langle \alpha^{k}\rangle_{C}$ is known as function of $k$, this can be promoted to a function of a real variable $p$ and inverted at $1$ to find the SAD exponent in the Griffiths phase in terms of the interaction distribution parameters. We expect this to hold when far from the vicinity of singularities such as the branch point of the Lambert W function in our outlined ER case.

\textit{Competitive interactions and instability}\textemdash To test robustness of these phenomena we relax the assumption of complete mutualism. We consider a fully connected case where the bulk of connections are given an intensive value $\hat{\mu}/S$.  ER structure is retained in reference to strong interactions of the same form giving an interaction distribution $P_{S}(\alpha) = (1-\frac{c}{S})\delta(\alpha-\frac{\hat{\mu}}{S}) + \frac{c}{
S}\delta(\alpha-\beta)$ \cite{supplementary}. For a mutualistic bulk interaction ($\hat{\mu}>0$) there is no qualitative difference with the system already considered; the unbounded growth phase expands as the SAD mean becomes $\frac{1}{1-(c\beta +|\hat{\mu}|)}$. Instead with competition ($\hat{\mu}<0$, see Fig. \ref{fig:pd comp}),  the mean abundance is $\langle x\rangle=\frac{1}{1-(c\beta-|\hat{\mu}|)}$ only for $c\beta<1$ for which there is no possibility of extinction. This extinction threshold is independent of $\hat{\mu}$ since extinctions are made possible not by the competition itself, but by competition involving species with large abundances. The unbounded growth boundary shifts as well due to the competition acting as regulation of the population abundance. 

These strongly interacting clusters can be understood from an ecological point of view as subcommunities of cooperative species existing together in mild competition, itself playing a role similar to the usual weak Gaussian disordered interactions. The superimposed mutualistic structure can significantly affect the properties of the SAD therefore even when the network is fully connected.

\begin{figure*}
\centering
\includegraphics[width=9cm]{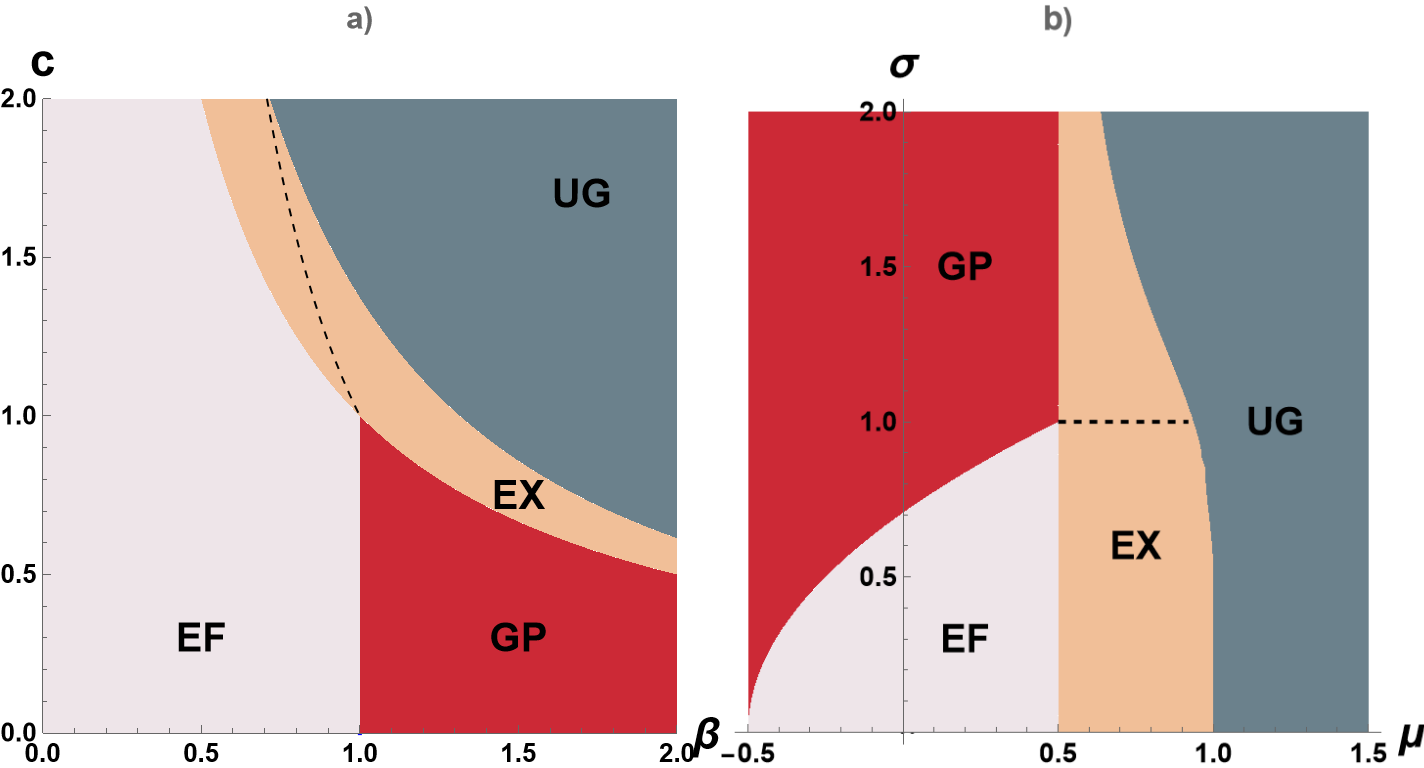}
\captionof{figure}{\label{fig:pd comp}Phase diagrams for the sparse GLV with background competition $\hat{\mu}=0.5$. a) is parametrised by $\beta, \ c$ of the ER network, while b) is parametrised by $\mu = c\beta-\hat{\mu}, \sigma= \sqrt{c}\beta$ for comparison with Gaussian phenomenology. The entire function (EF), Griffiths phase (GP) and unbounded growth (UG) phases and $\beta=1$, $c\beta=1$ boundaries remain from the $\hat{\mu}=0$ case in Fig. \ref{fig:pd}. A new phase where extinctions become possible (EX) is introduced however for $c\beta>1$ and before UG. Aside from a peak in $0$ taking account of these extinctions, simulations suggest the SAD transforms smoothly across the boundary from GP/EF. UG boundary is determined approximately due to the impact of extinctions. The dotted line represents $c\beta^2=1$, the earliest possible case in the EX phase where there are multiple attractors in the solution to the GLV and thus a fixed point is not guaranteed. } 
\end{figure*}

\textit{Conclusions}\textemdash In this Letter we have seen that in the paradigmatic random generalised Lotka-Volterra system, strong mutualistic interactions from a distribution with exponentially decaying tails can via dynamical evolution lead to a fixed point distribution with algebraically decaying tails of varying exponent. We argued that this phenomenon is comparable to the Griffiths phase and is present for different classes of models, even with different topology and under the introduction of competitive interactions. It would be interesting to explore the full implications of the extinctions that competition of this form and others \cite{marcusLocalExtensiveFluctuations2024} could have on the stationary and dynamic system properties. 

\bibliographystyle{unsrt}
\bibliography{bibs}

\begin{thebibliography}{10}

\bibitem{frankCommonPatternsNature2009}
Steven~A. Frank.
\newblock The {{Common Patterns}} of {{Nature}}.
\newblock {\em Journal of evolutionary biology}, 22(8):1563, June 2009.

\bibitem{guimaraesStructureEcologicalNetworks2020}
Paulo~R. Guimar{\~a}es.
\newblock The {{Structure}} of {{Ecological Networks Across Levels}} of {{Organization}}.
\newblock {\em Annual Review of Ecology, Evolution, and Systematics}, 51(Volume 51, 2020):433--460, November 2020.

\bibitem{seguraMetabolicBasisFat2019}
Angel~M. Segura and Gonzalo Perera.
\newblock The {{Metabolic Basis}} of {{Fat Tail Distributions}} in {{Populations}} and {{Community Fluctuations}}.
\newblock {\em Frontiers in Ecology and Evolution}, 7, May 2019.

\bibitem{ser-giacomiUbiquitousAbundanceDistribution2018}
Enrico {Ser-Giacomi}, Lucie Zinger, Shruti Malviya, Colomban De~Vargas, Eric Karsenti, Chris Bowler, and Silvia De~Monte.
\newblock Ubiquitous abundance distribution of non-dominant plankton across the global ocean.
\newblock {\em Nature Ecology \& Evolution}, 2(8):1243--1249, August 2018.

\bibitem{azaeleStatisticalMechanicsEcological2016a}
Sandro Azaele, Samir Suweis, Jacopo Grilli, Igor Volkov, Jayanth~R. Banavar, and Amos Maritan.
\newblock Statistical mechanics of ecological systems: {{Neutral}} theory and beyond.
\newblock {\em Reviews of Modern Physics}, 88(3):035003, July 2016.

\bibitem{grilliMacroecologicalLawsDescribe2020}
Jacopo Grilli.
\newblock Macroecological laws describe variation and diversity in microbial communities.
\newblock {\em Nature Communications}, 11(1):4743, September 2020.

\bibitem{gallaDynamicallyEvolvedCommunity2018}
Tobias Galla.
\newblock Dynamically evolved community size and stability of random {{Lotka-Volterra}} ecosystems(a).
\newblock {\em Europhysics Letters}, 123(4):48004, September 2018.

\bibitem{buninEcologicalCommunitiesLotkaVolterra2017}
Guy Bunin.
\newblock Ecological communities with {{Lotka-Volterra}} dynamics.
\newblock {\em Physical Review E}, 95(4):042414, April 2017.

\bibitem{baronBreakdownRandomMatrixUniversality2023a}
Joseph~W. Baron, Thomas~Jun Jewell, Christopher Ryder, and Tobias Galla.
\newblock Breakdown of {{Random-Matrix Universality}} in {{Persistent Lotka-Volterra Communities}}.
\newblock {\em Physical Review Letters}, 130(13):137401, March 2023.

\bibitem{newmanStructureRealworldNetworks2018}
Mark Newman.
\newblock The structure of real-world networks.
\newblock In Mark Newman, editor, {\em Networks}, page~0. Oxford University Press, July 2018.

\bibitem{busielloExplorabilityOriginNetwork2017}
Daniel~M. Busiello, Samir Suweis, Jorge Hidalgo, and Amos Maritan.
\newblock Explorability and the origin of network sparsity in living systems.
\newblock {\em Scientific Reports}, 7(1):12323, September 2017.

\bibitem{descheemaekerHeavytailedAbundanceDistributions2021}
Lana Descheemaeker, Jacopo Grilli, and Sophie {de Buyl}.
\newblock Heavy-tailed abundance distributions from stochastic {{Lotka-Volterra}} models.
\newblock {\em Physical Review E}, 104(3):034404, September 2021.

\bibitem{gibbs2023can}
Theo Gibbs, Gabriel Gellner, Simon~A Levin, Kevin~S McCann, Alan Hastings, and Jonathan~M Levine.
\newblock Can higher-order interactions resolve the species coexistence paradox?
\newblock {\em bioRxiv : the preprint server for biology}, pages 2023--06, 2023.

\bibitem{suweisGeneralizedLotkaVolterraSystems2024a}
Samir Suweis, Francesco Ferraro, Christian Grilletta, Sandro Azaele, and Amos Maritan.
\newblock Generalized {{Lotka-Volterra Systems}} with {{Time Correlated Stochastic Interactions}}.
\newblock {\em Physical Review Letters}, 133(16):167101, October 2024.

\bibitem{dedominicisDynamicsSubstituteReplicas1978a}
C.~De~Dominicis.
\newblock Dynamics as a substitute for replicas in systems with quenched random impurities.
\newblock {\em Physical Review B}, 18(9):4913--4919, November 1978.

\bibitem{aguirre-lopezHeterogeneousMeanfieldAnalysis2024a}
Fabi{\'a}n {Aguirre-L{\'o}pez}.
\newblock Heterogeneous mean-field analysis of the generalized {{Lotka}}--{{Volterra}} model on a network.
\newblock {\em Journal of Physics A: Mathematical and Theoretical}, 57(34):345002, August 2024.

\bibitem{metzDynamicalMeanFieldTheory2025}
Fernando~L. Metz.
\newblock Dynamical {{Mean-Field Theory}} of {{Complex Systems}} on {{Sparse Directed Networks}}.
\newblock {\em Physical Review Letters}, 134(3):037401, January 2025.

\bibitem{poleyInteractionNetworksPersistent2025}
Lyle Poley, Tobias Galla, and Joseph~W. Baron.
\newblock Interaction networks in persistent {{Lotka-Volterra}} communities.
\newblock {\em Physical Review E}, 111(1):014318, January 2025.

\bibitem{parkIncorporatingHeterogeneousInteractions2024d}
Jong~Il Park, Deok-Sun Lee, Sang~Hoon Lee, and Hye~Jin Park.
\newblock Incorporating {{Heterogeneous Interactions}} for {{Ecological Biodiversity}}.
\newblock {\em Physical Review Letters}, 133(19):198402, November 2024.

\bibitem{allesina2012stability}
Stefano Allesina and Si~Tang.
\newblock Stability criteria for complex ecosystems.
\newblock {\em Nature}, 483(7388):205--208, 2012.

\bibitem{khorunzhiyLifshitzTailsSpectra2006}
Oleksiy Khorunzhiy, Werner Kirsch, and Peter M{\"u}ller.
\newblock Lifshitz tails for spectra of {{Erd{\H o}s}}--{{R{\'e}nyi}} random graphs.
\newblock {\em The Annals of Applied Probability}, 16(1):295--309, February 2006.

\bibitem{cugliandoloMultifractalPhaseWeighted2024a}
Leticia~F. Cugliandolo, Gr{\'e}gory Schehr, Marco Tarzia, and Davide Venturelli.
\newblock Multifractal phase in the weighted adjacency matrices of random {{Erd{\H o}s}}--{{R{\'e}nyi}} graphs.
\newblock {\em Physical Review B}, 110(17):174202, November 2024.

\bibitem{fisherCriticalBehaviorRandom1995}
Daniel~S. Fisher.
\newblock Critical behavior of random transverse-field {{Ising}} spin chains.
\newblock {\em Physical Review B}, 51(10):6411--6461, March 1995.

\bibitem{tonoloGeneralizedLotkaVolterraModel2025}
Tommaso Tonolo, Maria~Chiara Angelini, Sandro Azaele, Amos Maritan, and Giacomo Gradenigo.
\newblock Generalized {{Lotka-Volterra}} model with sparse interactions: Non-{{Gaussian}} effects and topological multiple-equilibria phase, March 2025.

\bibitem{munozGriffithsPhasesComplex2010}
Miguel~A. Mu{\~n}oz, R{\'o}bert Juh{\'a}sz, Claudio Castellano, and G{\'e}za {\'O}dor.
\newblock Griffiths {{Phases}} on {{Complex Networks}}.
\newblock {\em Physical Review Letters}, 105(12):128701, September 2010.

\bibitem{morettiGriffithsPhasesStretching2013}
Paolo Moretti and Miguel~A. Mu{\~n}oz.
\newblock Griffiths phases and the stretching of criticality in brain networks.
\newblock {\em Nature Communications}, 4(1):2521, October 2013.

\bibitem{brayGriffithsSingularitiesRandom1989}
A.~J. Bray and Deng Huifang.
\newblock Griffiths singularities in random magnets: {{Results}} for a soluble model.
\newblock {\em Physical Review B}, 40(10):6980--6986, October 1989.

\bibitem{mallminChaoticTurnoverRare2024b}
Emil Mallmin, Arne Traulsen, and Silvia De~Monte.
\newblock Chaotic turnover of rare and abundant species in a strongly interacting model community.
\newblock {\em Proceedings of the National Academy of Sciences}, 121(11):e2312822121, March 2024.

\bibitem{marcusLocalCollectiveTransitions2022}
Stav Marcus, Ari~M. Turner, and Guy Bunin.
\newblock Local and collective transitions in sparsely-interacting ecological communities.
\newblock {\em PLOS Computational Biology}, 18(7):e1010274, July 2022.

\bibitem{marcusLocalExtensiveFluctuations2024}
Stav Marcus, Ari~M. Turner, and Guy Bunin.
\newblock Local and extensive fluctuations in sparsely interacting ecological communities.
\newblock {\em Physical Review E}, 109(6):064410, June 2024.

\bibitem{rohrWillLargeComplex2025a}
Rudolf~P. Rohr, Louis-F{\'e}lix Bersier, and Roger Arditi.
\newblock Will a large complex model ecosystem be viable? {{The}} essential role of positive interactions.
\newblock {\em Ecology}, 106(3):e70064, 2025.

\bibitem{haleMutualismIncreasesDiversity2020}
Kayla R.~S. Hale, Fernanda~S. Valdovinos, and Neo~D. Martinez.
\newblock Mutualism increases diversity, stability, and function of multiplex networks that integrate pollinators into food webs.
\newblock {\em Nature Communications}, 11(1):2182, May 2020.

\bibitem{griffithsNonanalyticBehaviorCritical1969}
Robert~B. Griffiths.
\newblock Nonanalytic {{Behavior Above}} the {{Critical Point}} in a {{Random Ising Ferromagnet}}.
\newblock {\em Physical Review Letters}, 23(1):17--19, July 1969.

\bibitem{stoneFeasibilityStabilityLarge2018a}
Lewi Stone.
\newblock The feasibility and stability of large complex biological networks: A random matrix approach.
\newblock {\em Scientific Reports}, 8:8246, 2018.

\bibitem{gohGlobalStabilityManySpecies1977}
B.~S. Goh.
\newblock Global {{Stability}} in {{Many-Species Systems}}.
\newblock {\em The American Naturalist}, 111(977):135--143, 1977.

\bibitem{akjouj2024complex}
Imane Akjouj, Matthieu Barbier, Maxime Clenet, Walid Hachem, Myl{\`e}ne Ma{\"\i}da, François Massol, Jamal Najim, and Viet~Chi Tran.
\newblock Complex systems in ecology: A guided tour with large {{Lotka}}–{{Volterra}} models and random matrices.
\newblock {\em Proceedings of the Royal Society A}, 480(2285):20230284, 2024.

\bibitem{flajoletAnalyticCombinatorics2009}
Philippe Flajolet and Robert Sedgewick.
\newblock {\em Analytic {{Combinatorics}}}.
\newblock Cambridge University Press, 2009.

\bibitem{flajoletSingularityAnalysisGenerating1990}
Philippe Flajolet and Andrew Odlyzko.
\newblock Singularity {{Analysis}} of {{Generating Functions}}.
\newblock {\em SIAM Journal on Discrete Mathematics}, 3(2):216--240, May 1990.

\bibitem{supplementary}
See {{Supplemental Material}} at [] for detailed derivations of {{CGF}} and {{SAD}} expressions.

\bibitem{azaeleGeneralizedDynamicalMean2024}
Sandro Azaele and Amos Maritan.
\newblock Generalized {{Dynamical Mean Field Theory}} for {{Non-Gaussian Interactions}}.
\newblock {\em Physical Review Letters}, 133(12):127401, September 2024.

\bibitem{brayDiffusionSparselyConnected1988}
A.~J. Bray and G.~J. Rodgers.
\newblock Diffusion in a sparsely connected space: {{A}} model for glassy relaxation.
\newblock {\em Physical Review B}, 38(16):11461--11470, December 1988.

\bibitem{rodgersDensityStatesSparse1988}
G.~J. Rodgers and A.~J. Bray.
\newblock Density of states of a sparse random matrix.
\newblock {\em Physical Review B}, 37(7):3557--3562, March 1988.

\bibitem{lukacsCharacteristicFunctions1970}
Eugene Lukacs.
\newblock {\em Characteristic Functions.}
\newblock Griffin, London, 2nd ed., revised \& enlarged. edition, 1970.

\bibitem{mandelbrotFractalsFormChance1977}
Benoit~B. Mandelbrot.
\newblock {\em {Fractals: form, chance, and dimension}}.
\newblock W. H. Freeman, San Francisco, 1977.

\bibitem{jackiwHowSuperrenormalizableInteractions1981}
R.~Jackiw and S.~Templeton.
\newblock How super-renormalizable interactions cure their infrared divergences.
\newblock {\em Physical Review D}, 23(10):2291--2304, May 1981.

\bibitem{vojtaRareRegionEffects2006}
Thomas Vojta.
\newblock Rare region effects at classical, quantum and nonequilibrium phase transitions.
\newblock {\em Journal of Physics A: Mathematical and General}, 39(22):R143, May 2006.

\bibitem{brayNatureGriffithsPhase1987}
A.~J. Bray.
\newblock Nature of the {{Griffiths}} phase.
\newblock {\em Physical Review Letters}, 59(5):586--589, August 1987.

\bibitem{odorLocalizationTransitionLifschitz2014}
G{\'e}za {\'O}dor.
\newblock Localization transition, {{Lifschitz}} tails and rare-region effects in network models.
\newblock {\em Physical Review E}, 90(3):032110, September 2014.

\bibitem{munozColloquiumCriticalityDynamical2018}
Miguel~A. Mu{\~n}oz.
\newblock Colloquium: {{Criticality}} and dynamical scaling in living systems.
\newblock {\em Reviews of Modern Physics}, 90(3):031001, July 2018.

\bibitem{grinsteinGenericScaleInvariance1991}
G.~Grinstein.
\newblock Generic scale invariance in classical nonequilibrium systems (invited).
\newblock {\em Journal of Applied Physics}, 69(8):5441--5446, April 1991.

\bibitem{imryGriffithsSingularityFinite1977}
Yoseph Imry.
\newblock Griffiths singularity in finite macroscopically large dilute {{Ising}} models.
\newblock {\em Physical Review B}, 15(9):4448--4450, May 1977.

\bibitem{breimanProbability1968}
Leo Breiman.
\newblock {\em Probability}.
\newblock SIAM, January 1968.

\bibitem{alonProbabilisticMethod2016}
Noga Alon and Joel~H. Spencer.
\newblock {\em The {{Probabilistic Method}}}.
\newblock Wiley Publishing, 4th edition, 2016.

\bibitem{corlessLambertWFunction1996}
R.~M. Corless, G.~H. Gonnet, D.~E.~G. Hare, D.~J. Jeffrey, and D.~E. Knuth.
\newblock On the {{LambertW}} function.
\newblock {\em Advances in Computational Mathematics}, 5(1):329--359, December 1996.

\end{thebibliography}

\onecolumn
\clearpage

\renewcommand{\thesection}{S\arabic{section}}
\renewcommand{\theequation}{S\arabic{equation}}
\renewcommand{\thefigure}{S\arabic{figure}}
\renewcommand{\thetable}{S\arabic{table}}

\setcounter{section}{0}
\setcounter{equation}{0}
\setcounter{figure}{0}
\setcounter{table}{0}

\title{Strong disorder in GLV Supplementary}
\author{Tommaso Jack Leonardi
  $^{1}$, Amos Maritan$^{1}$, Sandro Azaele$^{1}$\\
\emph{$^{1}$ Dipartimento di Fisica “G. Galilei”, Università di Padova, Italy } \\
 \texttt{tommasojack.leonardi@phd.unipd.it}
}

\maketitle
\tableofcontents
\newpage

\section{Appendix A: Derive fixed point generating function equation}\label{sec: deriv}
\subsection{Cavity-style fixed point derivation}

We state the DMFT fixed point effective equation for generalised Lotka-Volterra dynamics \cite{azaeleGeneralizedDynamicalMean2024}, with fixed point abundance $x^{*}$ as a function of a self-consistently determined local field $\eta$ which represents the bath of the other species.

\begin{equation}\label{GLV fp}
    x^{*}(\eta^{*}) = (1 +\eta^{*})\Theta(1 +\eta^{*})
\end{equation}

We can quickly derive the self-consistency by considering the local field $\eta_{0}$ felt by an invading species ($x_{0}$), given by $\eta_{0}=\sum_{j=1}\alpha_{0j}x^{*}_{j/0}$ and its CGF in a cavity-like heuristic. The notation $x^{*}_{j/0}$ refers to the pre-existing community at a presumed fixed point, the values of which are of course independent of the interaction parameters $\alpha_{0j}$ with the invading species. This allows the averages to be performed exactly, but since each species can be treated as the invading species, this gives the local field felt by each species up to $\mathcal{O}(1/S)$ error.

\begin{equation}\label{eq:stat supp}
    \begin{split}
         F_{\eta}(z) &= \ln{\langle \exp{(-i\eta z )}\rangle_{\eta}}\\&= \ln{\langle \exp{(-iz\sum_{j=1}^{S}\alpha_{0j}x^{*}_{j/0})\rangle_{\alpha}}} = \ln{\exp{\frac{1}{S}\sum_{j=1}^{S}F_{\alpha}(x^{*}_{j/0}z)}}\\
         &=\frac{1}{S}\sum_{j=1}^{S}F_{\alpha}(x^{*}_{j/0}z) \ \overset{S\to\infty}{=} \ \langle F_{\alpha}(x z)\rangle_{x}
    \end{split}
\end{equation}
where $\langle \rangle_{x}$ represents an average over the fixed point abundance distribution. Due to Eq. \eqref{GLV fp} this distribution is directly linked to the distribution of $\eta$ itself; hence the self-consistency.
From this we have the equivalent relations for the cumulants of $\eta$.
\begin{equation}
    \langle\eta^{*r}\rangle_{C} = \langle \alpha^{r}\rangle_{C} \langle x^{*r}\rangle_{\eta}\\
\end{equation}

We note that it must be that $\langle x^{*r}\rangle_{\eta}\geq0$ as they are moments. Thus, in the case that  $\langle \alpha^{r}\rangle_{C}\geq0$, automatically also $\langle\eta^{*r}\rangle_{C}\ge0$.

From the distribution of the noise at the fixed point we can find the distribution of $x^{*}$, $Q_{x}(x)$, in ecological contexts known as the species abundance distribution (SAD). We insert the solution of Equation \ref{GLV fp}

\begin{equation}
\begin{split}
    P_{x}(x) \equiv \langle \delta(x-x^{*}(\eta))\rangle_{\eta} = \delta(x)\int_{-\infty}^{-1}d\eta P_{\eta}(\eta) + P_{\eta}(x-1)\Theta(x)
    \end{split}
\end{equation}

Here is made explicit the separation between extinct species, a $\delta$ peak at $0$, with the distribution for the rest of the species related to the distribution of the noise by the inverse of $x^{*}(\eta)$. In the mutualistic case where there are no extinctions, we have simply $P_x(x) = P_{\eta}(x-1)$.

\subsection{ER single weight functional equation derivation}

The scaled CGF of the single weight ER interaction matrix is $F_{\alpha}(z) = c(e^{-i\beta z }-1)$. Using this in the equation for the local field CGF we get

\begin{equation}\label{eq:stat Poisson}
    \begin{split}
         F_{\eta}(z) &= \langle F_{\alpha}(x z)\rangle_{x}\\
         &= c\langle e^{-i\beta z}\rangle_{x}-c\\
         &= c\phi_{x}(\beta z) -c
    \end{split}
\end{equation}

As mentioned in the main text, since there are no extinctions, the characteristic function of the abundances can be related directly to that for the local field as $\phi_{x}(z)= e^{-iz}\phi_{\eta}(z)\equiv e^{-iz+F_{\eta}(z)}$. Thus we arrive at

\begin{equation}\label{eq:poisson functional equation}
    F_{\eta}(z) = c e^{-i\beta z+F_{\eta}(\beta z)} -c
\end{equation}

\section{Appendix B: $\beta=1$ derivation of Borel distribution}

Rearranging Eq. \eqref{eq:poisson functional equation} at $\beta=1$ gives the exact form of $F_{\eta}(z)$ for this case, which we can invert to find the probability distribution $P_{\eta}(\eta)$ and subsequently the SAD. 

\begin{equation}\label{eq:one supp}
    \begin{split}
        F_{\eta}(z)&= -c - W(-ce^{-c-iz})\\
        P_{\eta}(\eta) &= \int_{-\infty}^{\infty}\frac{dz}{2\pi}e^{i\eta z}\exp{\left(-c - W(-ce^{-c-iz})\right)} \\
        &= -\frac{1}{c}\int_{-\infty}^{\infty}\frac{dz}{2\pi}e^{i(\eta +1)z} W(-ce^{-c-iz})
    \end{split}
\end{equation}

We make a substitution $r=-ce^{-c-iz}, dr= -irdz$ to integrate on the circle $|r| = ce^{-c}$ on the complex plane (since $F_{\eta}(z)$ is periodic we can consider a single period on the real line of $z$) 

\begin{equation}
    \begin{split}
        P_{\eta}(\eta) &=  \frac{1}{c}\oint_{|r| = ce^{-c}}\frac{dr}{2\pi i }r^{-1}\left(\frac{-r}{ce^{-c}}\right)^{-(1+\eta)} W(r)\\
        &= \frac{(-1)^{-(1+\eta)}}{2\pi i c} (ce^{-c})^{1+\eta}\oint_{|r| = ce^{-c}}dr \ r^{-(2+\eta)} W(r)
    \end{split}
\end{equation}

Since $W(r)$ is analytic for $|r|<e^{-1}$, which is satisfied on the circle $|r| = ce^{-c}$ for $0<c<1$, we can use its Taylor expansion around the origin. This is $W(r) = \sum_{n=1}^{\infty}\frac{(-n)^{n-1}}{n!}r^{n}$. This allows us to easily use the residue theorem, as the simple pole when the exponent of $r$ is $-1$ will give the only contribution. This picks out the coefficient when $n=1+\eta$, i.e. $(-1)^{\eta}\frac{(1+\eta)^{\eta}}{(1+\eta )!}$, which also constrains $\eta$ to being 0 or a positive integer. Thus the normalised distribution of $\eta$

\begin{equation}
    \begin{split}
        P_{\eta}(\eta)= \frac{(c(1+\eta))^{\eta}e^{-c(1+\eta)}}{ (1+\eta )!} \ \ \ \ (n\in \mathbb{N})
    \end{split}
\end{equation}

giving the SAD directly since for feasible fixed points $P_{x}(x)= P_{\eta}(x-1)$

\begin{equation}
    \begin{split}
        P_{x}(x)= \frac{(cx)^{x-1}e^{-cx}}{ x!} \ \ \ \ (x\in \mathbb{N}\textbackslash0)
    \end{split}
\end{equation}

 the Borel distribution.

The Stirling approximation for $x!$ can be used to see behaviour for very large $x$ which can be approximated as a continuous variable. 
$x! \sim x^{x+\frac{1}{2}}e^{-x}$ gives

\begin{equation}
    \begin{split}
        P_{x}(x)\sim \frac{x^{-\frac{3}{2}}e^{-x(c-1-\ln{c})} }{c}
    \end{split}
\end{equation}
This same exponent is in fact also shared for generic pointwise local functional equations for $F_{\eta}(z)$ \cite{flajoletAnalyticCombinatorics2009}.

The principal branch of the Lambert W function has a branch point when its argument is equal to $-e^{-1}$ \cite{corlessLambertWFunction1996}. In eq. \eqref{eq:one supp} this occurs for $z=ir=i(c-1-\log{c})$, giving the exponential cutoff in the large abundance tail. This implies that in $F_{\eta}(z,\beta=1)$ there is the branch cut $\{iy: y> r\}$ on the imaginary axis.  The large $x$ behaviour of the SAD will be dominated by the behaviour of the CGF near $z=0$. The singularity is thus most dominant in the SAD when $i r$ is sufficiently close to $0$, i.e. this behaviour, and thus the topological signal in the abundance distribution becomes more important as $c\to1$ even for $\beta\neq1$. 

\section{Appendix C: $\beta>1$}

\subsection{Full solution}

The full solution has the form of a continued exponential

\begin{equation}
    \begin{split}
        F_{\eta}(z)= -c + c\exp{(-c-i\beta z +c\exp{(-c-i\beta^2 z + c\exp(-c -i\beta^3 z + \dots)})}
    \end{split}
\end{equation}

By defining $\psi(z) = -c\exp{(-c-iz)}$ we can see the repeating structure more clearly

\begin{equation}
    \begin{split}
        F_{\eta}(z)= -c -\psi(\beta z)e^{-\psi(\beta^2 z)e^{-\psi(\beta^3 z)e^{\dots}}}
    \end{split}
\end{equation}

Truncating this expression gives good approximation around $z\to -i \infty$, however these expressions would not be normalised and in general are slow to converge around $z=0$, which is the point of interest for extraction of statistical behaviour. Thus we can rewrite this continued exponential in terms of Lambert W functions

\begin{equation}
    \begin{split}
        F_{\eta}(z)= -c -W(\psi(\beta z))\exp{\left[W(\psi(\beta z))-W(\psi(\beta^{2} z))\exp{\left[W(\psi(\beta ^2 z))-W(\psi(\beta^{3} z))\exp{\dots}\right]}\right]}
    \end{split}
\end{equation}

Truncation of this tower allows the tower of exponentials to remain infinite, while still being a finite expression. This will always pass through $F(z=0)=0$ and so is normalised, though its behaviour deviates from the true behaviour between $z=0$ and $z=-\infty$. Concisely we can collapse all but the last Lambert W function in a truncation up to order $\beta^r$ as 

\begin{equation}
    \begin{split}
        F_{\eta}(z)= -c -\psi(\beta z)e^{-\psi(\beta^2 z)e^{\dots e^{-W(\psi(\beta^{r}z))}}}
    \end{split}
\end{equation}

It is this form used in Fig.$2 \text{vi)}$ of the main text, with $r=4$. This is formally undefined in the upper half plane, which does not exclude its treatment as a CGF. For example, CGFs are known to be undefined on a half-plane for heavy-tailed distributions such as Pareto distributions or the lognormal distribution \cite[see Section 7]{lukacsCharacteristicFunctions1970}.

\subsection{Series approximation}\label{sec: series solution}

As reported in the main text, considering a regime where the CGF is small we can find the exact solution to the simpler problem below

\begin{equation}
    F_{\eta}(z) = c e^{-i\beta z}(1+F_\eta(\beta z))-c
\end{equation}

Making this approximation is equivalent to saying the characteristic function is related to the CGF as $\phi_{\eta}(z) \sim 1+ F_\eta(z)$ and so has the same functional dependence on $z$.

This has exact solution 

\begin{equation}\label{eq:series sol supp}
    F_{\eta}(z) = -c + (1-c)\sum_{n=1}^{\infty}c^n\exp{(-iz \sum_{j=1}^{n}\beta^{j})}
\end{equation}

This can be considered as the first in a series of asymptotic approximations to the complete solution to Equation \eqref{eq:poisson functional equation} for $c<1$. We see this by find the second order approximation as 

\begin{equation}
    \begin{split}
        F_{\eta}(z) = -c + \sum_{n=1}^{\infty}g_{n}\exp{(-iz \sum_{j=1}^{n}\beta^{j})}+\sum_{n_{1}=2}^{\infty}\sum_{n_{2}=2}^{\infty}\sum_{m=2}^{n_{2}}h_{n_{1},n_{2},m}\exp{(-iz\left( \sum_{j=1}^{n_{1}}\beta^{j}+ \sum_{j=m}^{n_{2}}\beta^{j}\right))}
    \end{split}
\end{equation}

The coefficients as set by using Equation \eqref{eq:poisson functional equation} and comparing coefficients at this order. This gives the following :

\begin{equation}
    \begin{split}
        g_{1} &= c(1-c+\frac{c^{2}}{2})\\
        n>1: g_{n} &=g_{n-1}c(1-c)= g_{1}(c(1-c))^{n-1}\\
        \quad\\
        h_{n_{1},n_{2},2} &=\frac{c}{2}g_{n_{1}-1}g_{n_{2}-1}= \frac{g_{1}^{2}}{2}c^{n_{1}+n_{2}-3}(1-c)^{n_{1}+n_{2}-4}\\
        m>2: h_{n_{1},n_{2},m} &=h_{n_{1},n_{2},2}(c(1-c))^{m-2}
    \end{split}
\end{equation}

The corrections to Equation \eqref{eq:series sol supp} appear in each term of the sum. However each of these second order corrections are of $\mathcal{O}(c)$ with respect to each term in the first order sum, and so on for each successive correction. For $c<1$ this is sufficient to treat Equation \eqref{eq:series sol supp} as a first approximation solution to the full functional equation \eqref{eq:poisson functional equation}.

The density distribution $P_{\eta}(\eta)$ under the first distribution is found by the reverse Fourier transform of the characteristic function $\phi_{\eta}(z) \approx 1+ F_{\eta}(z)$ with the approximation as above. This gives

\begin{equation}
    \begin{split}
        P_{\eta}(\eta)&= \int_{-\infty}^{\infty}\frac{dz}{2\pi} e^{i \eta z}\phi_{\eta}(z)\\
        &= (1-c)\int_{-\infty}^{\infty}\frac{dz}{2\pi} e^{i \eta z} +(1-c)\sum_{n=1}^{\infty}c^{n}\int_{-\infty}^{\infty}\frac{dz}{2\pi} e^{i \eta z}\exp{(-iz\sum_{j=1}^{n}\beta^{j})}\\
        &= (1-c)\delta(\eta) +(1-c)\sum_{n=1}^{\infty}c^{n}\delta(\eta-\sum_{j=1}^{n}\beta^{j})
    \end{split}
\end{equation}

At this point it is in fact simpler to consider the cumulative frequency distribution function as in \cite{mandelbrotFractalsFormChance1977} p.328

\begin{equation}
    \begin{split}
        \text{CDF}_{\eta}(\eta) &= \int_{\eta}^{\infty}d\eta' P_{\eta}(\eta')\\
        &= (1-c) \sum_{n=1}^{\infty}c^{n}\int_{\eta}^{\infty}d\eta' \delta(\eta'-\sum_{j=1}^{n}\beta^{j})
    \end{split}
\end{equation}
The sum will include terms only from a certain $n_0(\eta)$ corresponding to when $\sum_{j=1}^{n_{0}}\beta^{j} = \eta $ i.e. $n_{0} = \frac{\ln{(1+\frac{\beta-1}{\beta}\eta})}{\ln{\beta}}$

\begin{equation}
    \begin{split}
        \text{CDF}_{\eta}(\eta) &= (1-c)c^{n_{0}}\sum_{n=1}^{\infty}c^{n}\\
        &= c^{\frac{\ln{(1+\frac{\beta-1}{\beta}\eta})}{\ln{\beta}}}\\
        &= (1 + \frac{\beta-1}{\beta}\eta)^{\frac{\ln{c}}{\ln{\beta}}} 
    \end{split}
\end{equation}

For the density distribution of $\eta$ and thus also abundance $x$ this implies an overall exponent of $-(1-\frac{\ln{c}}{\ln{\beta}})$.

\subsection{Expansions around 0}\label{sec: expansions around 0}

Since the behaviour around $0$ is difficult to extract directly from the continued exponential, we return to the functional equation. The choice of expansion is highly dependent on the relative values of $c$ and $\beta$. Illustratively we consider the albeit unphysical case $\beta=1/c$.

\begin{equation}
    \begin{split}
        F(z,1/c) = c(\exp{(-z/c+F(z/c,1/c)}-1)
    \end{split}
\end{equation}

Taking ourselves to be near $z=0$ we use the ansatz $F(z,1/c) = a_{0} z \ln{z} +a_{1}z+\dots$. This gives

\begin{equation}
\begin{split}
    F(z,1/c) &= c\exp{(z/c +F(z/c,1/c)}-c\\
    a_{0} z \ln{z}+a_{1}z+O(z^{2}\ln{z})\dots&= c\exp{(z/c +a_{0}z/c \ln{z} -a_{0}/c z\ln{c} +a_{1}\frac{z}{c}+ \dots )}-c\\
    &= a_{0}z\ln{z}+z(a_{1}+1-a_{0}\ln{c})
\end{split}
\end{equation}

We can see that it is consistent to introduce the $z\ln{z}$ term and that its coefficient is set by comparing the linear $z$ term in $F$. So $a_{0}=\frac{1}{\ln{c}}$. This leaves $a_{1}$ undetermined as it is present equally on both sides.

Including the next orders where consistency follows we have

\begin{equation}
    F(z,1/c) = \frac{z\ln{z}}{\ln{c}}+a_{1}z +\frac{z^{2}(\ln{z})^{2}}{2(c-1)(\ln{c})^{2}}+  \frac{a_{1}(c-1)-1}{(c-1)^{2}\ln{c}}z^{2}\ln{z}+\frac{c+1+a_{1}(c-1)(a_{1}(c-1)-2)}{2(c-1)^{3}}z^{2}+\dots
\end{equation}

While at first order the presence of the exponential was not incorporated, at higher orders naturally there are contributions which must be taken into account self-consistently. Through this process coefficients can be found, though there remains throughout the indeterminacy of $a_{1}$, a degree of freedom in this local expansion. We discuss this in the next section.

We can extend this reasoning for all 'critical' $\beta$, i.e. the $\beta_{k}$ for which the $k$th interaction cumulant $\langle \alpha ^{k}\rangle_{C}$ is 1 ($k$ integer). We know from the cumulant definition that $\beta_{k} = c^{-\frac{1}{k}}$. 

We introduce the complete exponential Bell polynomials as the coefficients of the partitions of $n$ objects with partitions of size $j$ weighted by the corresponding element of series $\{x_{j}\}$.

\begin{equation}
    B_{n}(x_{1},\dots,x_{n}) = \left(\frac{\partial}{\partial t}\right)^{n}\exp{\left(\sum_{j=1}^{n}x_{j}\frac{t^{j}}{j!}\right)\bigg\rvert_{t=0}}  = n! \sum_{\{\textbf{j}\}, \ \sum_{i=1}^{n}ij_{i}=n} \ \prod_{i=1}^{n}\frac{x_{i}^{j_{i}}}{(i!)^{j_{i}}j_{i}!}
\end{equation}

In general for a probability distribution with moments $\mu_{i}, i\in[1,n]$ and cumulants $\kappa_{i}, i\in[1,n]$ we have for $i \in[1,n]$

\begin{equation}
    \mu_{i} = B_{i}(\kappa_{1},\dots,\kappa_{i})
\end{equation}

For $\beta=\beta_{k}$ we take $F(z,\beta_{k}) = g(z) + a_{0,k}z^{k}\ln{z}+a_{1,k}z^{k}+\mathcal{O}(z^{k+1})$ for an analytic function $g(z)$. 

Expanding the functional equation around $z=0$ retaining terms in $g(z)$ up to order $z^{k}$ we have

\begin{equation}
\begin{split}
    g(z) + a_{0,k}z^{k}\ln{z}+ a_{1,k}z^{k}&= c\exp{(\beta_{k}z+g(\beta_{k}z)+a_{0,k}\beta_{k}^{k}z^{k}\ln{\beta_{k}z}+a_{1,k}\beta_{k}^{k}z^{k})-c}\\
    \sum_{r=1}^{k-1}\frac{a_{r}z^{r}}{r!} + a_{0,k}z^{k}\ln{z}+ a_{1,k}z^{k} &= c\exp{(c^{-\frac{1}{k}}z+\sum_{l=1}^{k-1}\frac{a_{l}z^{l}}{l!})}(1+a_{0,k}z^{k}/c \ln{z} -a_{0,k}(\ln{c})z^{k}/c+a_{1,k}z^{k}/c)-c\\
    &= c\sum_{l=1}^{k-1}c^{-\frac{l}{k}}\frac{B_{l}(1+a_{1},\dots,a_{l})}{l!}z^{l} +a_{0}z^{k} \ln{z} -a_{0}\ln{c} \ z^{k}+a_{1,k}z^{k}
\end{split}
\end{equation}

From this we can compare coefficients at each order of $z$. For $r<k$ we have

\begin{equation}
    a_{r} = \frac{c^{1-r/k}}{1-c^{1-r/k}}B_{r}(1+a_{1},\dots,a_{r-1},0)
\end{equation}

Which we can see will get very large for $r=k-1$ and large $k$. These are in fact the cumulants of the noise distribution up to this order and we can relabel them $a_{r}=\langle\eta^{r}\rangle_{C}$ up to $r=k-1$.

Comparing $\mathcal{O}(z^{k}\ln{z})$ we see the term is consistent, and its coefficient is defined by comparing $\mathcal{O}(z^{k})$, and $a_{1,k}$ remains undetermined.

\begin{equation}
    a_{0,k} = \frac{B_{k}(1+a_{1},\dots,a_{k-1},0)}{\ln{c}}
\end{equation}

We selected the expansion for these values of $\beta$ wrt $c$ based on the relation $c \beta^{k}=1$, which gives when the $k$th cumulant of the noise distribution is expected to diverge. This justification via cumulants is agnostic on the behaviour for values of $\beta$ which are between $c^{-1/k}$ and $c^{-1/(k-1)}$ for integer $k$. However, directly from the functional equation we see that this relation remains key even in these intervals. Illustratively we consider when $\beta=c^{-1/p}$ for $1<p<2$, using the ansatz $F_{\eta}(z) = b_{1}z+b_{p}|z|^{p}+\dots$

\begin{equation}
    \begin{split}
        F_{\eta}(z,c^{-1/p}) &= c(\exp{(c^{-1/p}z+F_{\eta}(c^{-1/p}z,c^{-1/p}))}-1)\\
        &=c(\exp{(c^{-1/p}z+b_{1}c^{-1/p}z+b_{p}c^{-1}|z|^{p})}-1)\\
        &= c^{1-1/p}(1+b_{1})z + b_{p}|z|^{p}
        \end{split}
\end{equation}
We see that a term proportional to $|z|^{p}$ is consistent in the functional equation, though its coefficient cannot be determined by this expansion alone, as we have isolated the homogeneous regime. We write the successive terms for this case, using $\beta$ known to satisfy $c\beta^{p}=1$ 

\begin{equation}
    \begin{split}
        F_{\eta}(z,c^{-1/p}) = \frac{c\beta}{1-c\beta}z+ a_{p}|z|^{p}+\frac{c\beta^{2}}{2(1-c\beta)^{2}(1-c\beta^{2})}z^{2}+\frac{c\beta^{2} a_{p}}{(1-\beta)(1-c\beta)}|z|^{1+p}+\frac{a_{p}^{2}}{2(c-1)}|z|^{2p}+\dots
    \end{split}
\end{equation}

In a similar way as with $\beta=c^{-1/k}$ for integer $k$, the expansion up to order $|z|^{p}$ will be formed of analytic terms with coefficients that are interpretable as the cumulants of the noise distribution. At the non-analyticity and beyond successive terms and coefficients can be found gradually, excepting a degree of freedom, which we leave in terms of $a_{p}$.

\begin{equation}
    \begin{split}
        F_\eta(z, \beta= c^{-1/p}) = \sum_{r=1}^{\lfloor{p}\rfloor}\frac{\langle \eta ^{r}\rangle_{C} }{r!}z^{r} + a_{p} |z|^{p} + \mathcal{O}(z^{\lceil p\rceil})
    \end{split}
\end{equation}
In summary we have

When $c \beta^{k} =1 $ for $ k \in \mathbb{N}$, in the limit $z\to 0$
\begin{equation}\label{eq: disc exp}
    \begin{split}
        F_\eta(z, \beta= c^{-1/k}) &= \sum_{r=1}^{k-1}\frac{\langle \eta ^{r}\rangle_{C} }{r!}(-iz)^{r} + a_{0,k} (-iz)^{k} \ln{|z|} + a_{1,k} z^{k} + \mathcal{O}(\max{(z^{2k} (\log{z})^{2k}, z^{k+1}\log{z})})\\
         a_{0,k}&= \frac{B_{k}(\langle \eta \rangle_{C},\dots,\langle \eta ^{k-1}\rangle_{C},0)}{\log{c}} 
    \end{split}
\end{equation}
When $c \beta^{p} =1 $ for $ p \in \mathbb{R}^{+}-\mathbb{N}$ , in the limit $z\to 0$

\begin{equation}\label{eq: cont exp}
    \begin{split}
        F_\eta(z, \beta= c^{-1/p}) = \sum_{r=1}^{\lfloor{p}\rfloor}\frac{\langle \eta ^{r}\rangle_{C} }{r!}(-iz)^{r} + a_{p} |z|^{p} + \mathcal{O}(z^{\lceil p\rceil})
    \end{split}
\end{equation}

While $a_{0,k}$ can be found for all integer $k$ in terms of the lower order cumulants, the coefficients $a_{1,k}, \ a_{p}$ cannot be locally determined. All the higher order terms' coefficients can be found in terms of this single free parameter ($a_{1,k}$ in the integer case, $a_{p}$ for the non-integer). This final parameter is found globally, similarly to how the Euler-Mascheroni constant appears in expansions of the exponential integral.

\subsection{Undetermined coefficients}\label{sec: undet coeff}

Alongside both types of non-analyticity we see in the expansions around $0$ there is a degree of freedom we are unable to determine locally. This can be understood as a consequence of the homogeneity of $F_{\eta}(z)$ at a certain scale of $z$ depending on the parameters $c,\beta$. This means the functional equation acts like the homogeneous equation $F_{\eta}(z)\approx cF_{\eta}(\beta z)$ which inherently cannot fix a coefficient for this component of the behaviour by itself. 

A bookwork example of this situation is the expansion of the exponential integral around 0 for which the Euler-Mascheroni constant is present.
Taking $E_{1}(x)=\int_x^{\infty}dt \frac{e^{-t}}{t}$, differentiating with respect to $x$, and integrating term by term of a Maclaurin expansion will result in the correct asymptotic expansion. However the integration constant cannot be determined by this local expansion and requires injection of the global information. Indeed the constant is the Euler-Mascheroni constant $\gamma$
\begin{equation}
    \begin{split}
        \gamma = \lim_{x\to0}\left(E_{1}(x)+\ln{x}\right)
    \end{split}
\end{equation}

The Euler-Mascheroni constant can be expressed as the limiting difference between the logarithm and a great deal of series, most famously the harmonic series. 

In the same way we can set the coefficients which from a simple expansions around $0$ of the Poisson functional equation through its deviation from the more significant terms. 
For the integer case and non-integer cases respectively
\begin{equation}
    \begin{split}
        a_{1,k} = \lim_{z\to 0 } \frac{1}{z^{k}}\left( F_\eta(z, \beta= c^{-1/k}) - \sum_{r=1}^{k-1}\frac{\langle \eta ^{r}\rangle_{C} }{r!}(-iz)^{r} - a_{0,k} (-iz)^{k} \log{|z|}\right)
    \end{split}
\end{equation}

\begin{equation}
    \begin{split}
        a_{p} = \lim_{z\to 0 } \frac{1}{|z|^{p}}\left( F_\eta(z, \beta= c^{-1/p}) - \sum_{r=1}^{\lfloor{p}\rfloor}\frac{\langle \eta ^{r}\rangle_{C} }{r!}(-iz)^{r}\right)
    \end{split}
\end{equation}

In particular we state the illustrative cases $k=1$ and $1<p<2$ for which we show the numerical extraction in a log-log plot for which the coefficient is $-\exp{y_{0}}$ for y-intercept $y_{0}$.
.
\begin{equation}\label{eq: coeff 2}
    \begin{split}
        a_{1,1} =  \lim_{z\to 0 } \frac{1}{z}\left( F_\eta(z, \beta= c^{-1}) - a_{0,1} z \log{|z|}\right)
    \end{split}
\end{equation}

\begin{equation}\label{eq: coeff 1.5}
    \begin{split}
        a_{p} =  \lim_{z\to 0 } \frac{1}{|z|^{p}}\left( F_\eta(z, \beta= c^{-1/p}) - \langle\eta\rangle z\right)
    \end{split}
\end{equation}

It would also be possible to use the series approximation Eq. \eqref{eq:series sol supp} to make analytical bounds on these coefficients.

\begin{figure*}
\includegraphics[width=17cm]{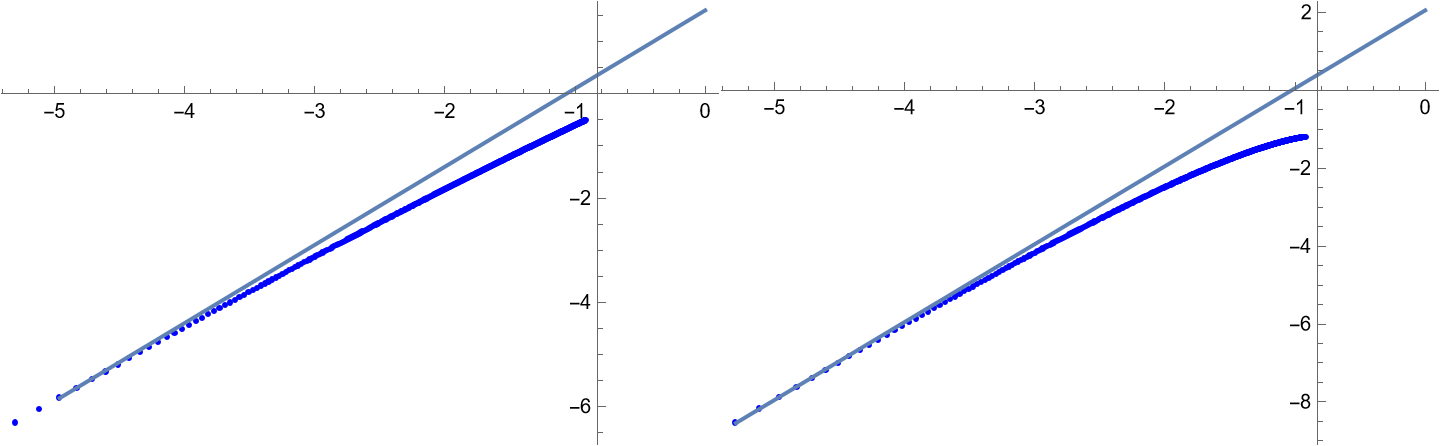}
\caption{\label{fig:numerical}Coefficients extracted as according to the text from the numerical solution to the functional equation, by looking at the y-intercept of the linear component on the log-log plot. (left) corresponds to Equation \eqref{eq: coeff 1.5} with $c=0.3 \ \beta=c^{-1/1.5}$ and (right) corresponds to Equation \eqref{eq: coeff 2} with $c=0.3 \ \beta=c^{-1/2}$}
\end{figure*}

\subsection{Retrieving the SAD}\label{sec: retrieve SAD}

Once we have an expression for the behaviour of the CGF of the coloured noise in a certain limit, we can translate this into that of the SAD distribution itself. In particular the behaviour of the CGF near 0 will give us the tails of the SAD as $x\to\infty$, with an appropriate transformation. 
The forms of non-analyticity we see are either $F_\eta(z,c^{-1/k})= f_{1}(z)+a_{0,k}z^{k}\ln{z}+\mathcal{O}(z^{k+1})$ for an integer $k$ or $F_\eta(z,c^{-1/p})= f_{2}(z)+a_{p}|z|^{p}+\mathcal{O}(z^{\lceil p \rceil})$ for non-integer $p$, where $f_{1},f_{2}$ are analytic.

The moment-generating function is $\phi_{\eta}(z,\beta)=\exp{F_{\eta}(z,\beta)}$ and if we retain to the same order then $\phi_\eta(z,c^{-1/k})= \tilde{f}_{1}(z)+a_{0,k}z^{k}\ln{z}+\mathcal{O}(z^{k+1})$ and $\phi_\eta(z,c^{-1/p})= \tilde{f}_{2}(z)+a_{p}|z|^{p}+\mathcal{O}(z^{\lceil p \rceil})$. We define new analytic functions $\tilde{f}_{1},\tilde{f}_{2}$ as the exponential introduces new terms.

The inverse Laplace transform of $\phi_{\eta}(z,c^{-1/k})$ is, as $\eta\to\infty$

\begin{equation}
    P_{\eta}(\eta)= \frac{k!(-1)^{k+1} a_{0,k}}{\eta^{k+1}}+\mathcal{O}\left(\frac{1}{\eta^{k+2}}\right)
\end{equation}

The higher order corrections arise both from the CGF expansion and the inverse Laplace transformation itself. Instead the inverse Laplace transformation of $\phi_\eta(z,c^{-1/p})$ is as $\eta\to\infty$

\begin{equation}
    P_{\eta}(\eta)= \frac{-a_{p}}{\eta^{p+1}\Gamma(-p)}+\mathcal{O}\left(\frac{1}{\eta^{p+2}}\right)
\end{equation}

We can see there is continuity of the exponent of the SAD as $p$ tends to an integer. The SAD itself can be retrieved with a simple translation due to the stationarity relation $x^{*}(\eta)=1+\eta$.

\begin{figure*}
\includegraphics[width=10cm]{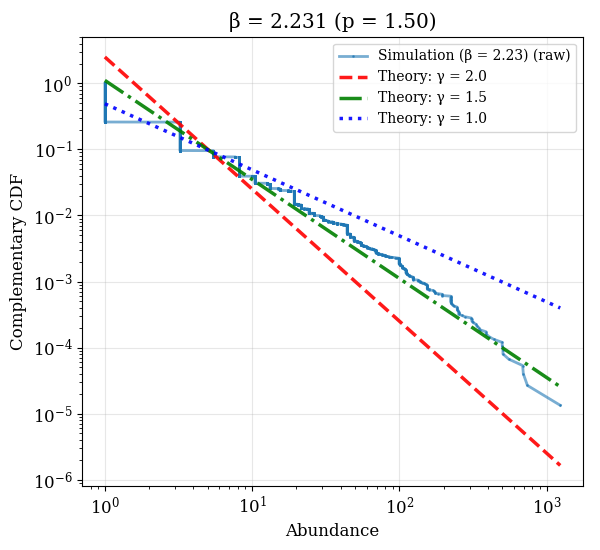}
\caption{\label{fig:powers}Predicted power law for the cumulative species abundance distributions compared with GLV simulation stationary abundances with 15 repetitions of systems with species number $S=5000$. Plotted is $c=0.3, \ \beta=c^{-1/1.5}$}
\end{figure*}
\section{Appendix D: Extensions}

\subsection{Asymptotics for ER and intensity distribution}\label{sec: ER and intensity dist}

\subsubsection{Two intensities}
\begin{equation}
    \begin{split}
        P_{S}(\alpha) = (1-\frac{c}{S})\delta(\alpha) + &\frac{c}{S}(q_{1}\delta(\alpha-\beta_{1})+q_{2}\delta(\alpha-\beta_{2}))\\
        q_{1}+q_{2}&=1
    \end{split}
\end{equation}

\begin{equation}
    \begin{split}
        F_{\alpha}(z)= c(q_{1}e^{-i\beta_{1}z}+q_{2}e^{-i\beta_{2}z}-1)
    \end{split}
\end{equation}

\begin{equation}
    \begin{split}
        F_{\eta}(z)= c(q_{1}e^{-i\beta_{1}z+F_{\eta}(\beta_{1}z)}+q_{2}e^{-i\beta_{2}z+F_{\eta}(\beta_{2}z)}-1)
    \end{split}
\end{equation}

With analytic expansion coefficients are

\begin{equation}
    \begin{split}
        \langle \eta^{r}\rangle_{C} &= \frac{\langle \alpha^{r}\rangle_{C}}{1-\langle \alpha^{r}\rangle_{C}}B_{r}(1+\langle \eta\rangle_{C},\langle \eta^{2}\rangle_{C},\dots,\langle \eta^{r-1}\rangle_{C},0) 
    \end{split}
\end{equation}

Here $\langle \alpha^{r}\rangle_{C} = c(q_{1}\beta_{1}^{r}+q_{2}\beta_{2}^{r})$.

This has similar non-analyticities to the single $\beta$ case. For example, when the first cumulant is 1 $cq_{1}\beta_{1}+cq_{2}\beta_{2}=1$ then near $z=0$ $F_{\eta}(z) = a_{0}z\ln{z}+a_{1} z +\dots$ where $a_{0}= -\frac{1}{cq_{1}\beta_{1}\ln\beta_{1}+cq_{2}\beta_{2}\ln\beta_{2}}$. 

In the same way as for single $\beta$, we have the expansions for the region of the phase diagram where at least one of $\beta_{i} >1$. 
When $c(q_{1}\beta_{1}^{k}+q_{2}\beta_{2}^{k}) =1 $ for $ k \in \mathbb{N}$, in the limit $z\to 0$

\begin{equation}
    \begin{split}
        F_\eta(z, k) &= \sum_{r=1}^{k-1}\frac{\langle \eta ^{r}\rangle_{C} }{r!}(-iz)^{r} + a_{0,k} (-iz)^{k} \log{|z|} + a_{1,k} z^{k} + \mathcal{O}(\max{(z^{2k} (\log{z})^{2k}, z^{k+1}\log{z})})\\
         a_{0,k}&= -\frac{B_{k}(\langle \eta \rangle_{C},\dots,\langle \eta ^{k-1}\rangle_{C},0)}{c q_{1}\beta_{1}^{k}\ln{\beta_{1}}+c q_{2}\beta_{2}^{k}\ln{\beta_{2}}} 
    \end{split}
\end{equation}

When $c(q_{1}\beta_{1}^{p}+q_{2}\beta_{2}^{p}) =1 $ for $ p \in \mathbb{R}^{+}-\mathbb{N}$ , in the limit $z\to 0$

\begin{equation}
    \begin{split}
        F_\eta(z, p) = \sum_{r=1}^{\lfloor{p}\rfloor}\frac{\langle \eta ^{r}\rangle_{C} }{r!}(-iz)^{r} + a_{p} |z|^{p} + \mathcal{O}(z^{\lceil p\rceil})
    \end{split}
\end{equation}

$a_{0,k}$ and $a_{1,k}$ are found globally in a similar way as with single $\beta$. 

\subsubsection{$n$ discrete intensities}
We can extend directly to $n$ different $\beta$ with the conditions $c\sum_{r=1}^{n}q_{r}\beta_{r}^{(k,p)}=1 $ for $k$ or $p$. Naturally we must have $\sum_{r=1}^{n}q_{r}=1$ for normalisation.

The functional equation becomes
\begin{equation}
    F_{\eta}(z)= c(\sum_{r=1}^{n}q_{r}e^{-i\beta_{r}z+F_{\eta}(\beta_{r}z)}-1)
\end{equation}

Then $a_{0,k} = -\frac{B_{k}(\langle \eta \rangle_{C},\dots,\langle \eta ^{k-1}\rangle_{C},0)}{c \sum_{r=1}^{n} q_{r}\beta_{r}^{k}\ln{\beta_{r}}} $. The expansions at a certain $p,k$ have the same form, though of course $\langle \eta ^{r}\rangle_{C}$ update accordingly as do the globally determined $a_{1,k}$ and $a_{p}$. 

\subsubsection{Continuum of intensities}\label{sec: int dist ER}
Promoting now to a distribution of $\beta$, the sum over discrete probabilities $q_{r}$ becomes the normalisation of the $\beta$ values $\int_{0}^{\infty}d\beta P_{\beta}(\beta) =1$. 

\begin{equation}
    P_{S}(\alpha)= (1-\frac{c}{S})\delta(\alpha) + \frac{c}{S}P_{\beta}(\alpha)
\end{equation}

We restrict the domain to the positive real line in the mutualistic case. As a compound poisson this would give an interaction distribution with cumulants $\langle \alpha^{k}\rangle_{C} = c\langle \beta^{k}\rangle$. This expression being equal to $1$ allows calculation of the exponent, and the same expansions conjecture can be used in the continuum limit. 
We restrict the domain to the positive real line in the mutualistic case. As a compound poisson this would give an interaction distribution with cumulants $\langle \alpha^{k}\rangle_{C} = c\langle \beta^{k}\rangle$. This expression being equal to $1$ allows calculation of the exponent, and the same expansions conjecture can be used in the continuum limit. 

The functional equation becomes (where we link with the derivation from the original generalised functional equation). 
\begin{equation}
    F_{\eta}(z)= c(\int_{0}^{\infty} d\beta \ P_{\beta}(\beta)e^{-i\beta z+F_{\eta}(\beta z)}-1) = c(\langle \phi_{\beta}(z(1+\eta))\rangle_{\eta}-1)
\end{equation}

$ a_{0,k} = -\frac{B_{k}(\langle \eta \rangle_{C},\dots,\langle \eta ^{k-1}\rangle_{C},0)}{c \langle \beta^{k} \ln{\beta}\rangle}$

The series solution valid for $c<1$ begins analogously to that of the single intensity case, again with error in the $n$th term of the sum being of order $\mathcal{O}(c^{n+1})$

\begin{equation}
    \begin{split}
        F_{\eta}(z)=-c + (1-c)\sum_{n=1}^{\infty}\int_{0}^{\infty}\prod_{i=1}^{n}[d\beta_{i}P_{\beta}(\beta_{i})] \ c^{n}\exp{(-iz\sum_{j=1}^{n}\prod_{i=1}^{j}\beta_{i})} +\dots
    \end{split}
\end{equation}
\subsection{Configuration model}

For another example of structure beyond ER, we consider an interaction matrix generated with configuration model structure where a degree distribution $p_{k}$ is specified for $k$, the degree per node. 

\begin{equation}
    P_{S}(\alpha_{ij}) = \left(1-\frac{k_{i}k_{j}}{\langle k\rangle S}\right)\delta(\alpha_{ij}) +\frac{k_{i}k_{j}}{\langle k\rangle S}P_{\beta}(\alpha_{ij})
\end{equation}
For this case it is interesting in particular to consider the subsystem of the GLV consisting of the species with a certain degree $k$ wlog

\begin{equation}
    \begin{split}
        \dot{x}^{(k_{i})}_{i} &= x^{(k_{i})}_{i}(1-x^{(k_{i})}_{i}+\sum_{j\neq i}^{S}\alpha_{ij}x^{(k_{j})}_{j})\\
    \end{split}
\end{equation}

The local field at the fixed point $\eta_{k_{i}}=\sum_{j\neq i}^{S}\alpha_{ij}x^{(k_{j})}_{j})$ has the CGF

\begin{equation}
    \begin{split}
        F_{\eta^{(k_{i}})}(z)&= \sum_{j\neq i}^{S}\ln{\langle \exp{\left(-iz\alpha_{ij}x^{(k_{j})}_{j}\right) }\rangle_{\alpha_{ij}}}\\
        &=\sum_{j\neq i}^{S} \ln{\left(1 - \frac{k_{i}k_{j}}{\langle k\rangle S}+ \frac{k_{i}k_{j}}{\langle k\rangle S}\phi_{\beta}(x^{(k_{j})}_jz)\right)}\\
        &= \frac{k_{i}}{\langle k\rangle} \frac{1}{S}\sum_{j\neq i}^{S} k_{j} \left(\phi_{\beta}(x^{(k_{j})}_jz)-1\right)\\
        &= \frac{k_{i}}{\langle k\rangle}\sum_{k'} k'\frac{S_{k'}}{S}\frac{1}{ S_{k'}}\sum_{j=1}^{S_{k'}}\left(\phi_{\beta}(x^{(k')}_jz)-1\right)\\
        &= \frac{k_{i}}{\langle k\rangle}\sum_{k'} k'p_{k'}\frac{1}{ S_{k'}}\sum_{j=1}^{S_{k'}}\left(\phi_{\beta}(x^{(k')}_jz)-1\right)\\
        F_{\eta^{(k)}}(z)&= \frac{k}{\langle k\rangle}\sum_{k'} k'p_{k'}\langle\phi_{\beta}(x^{(k')}z)-1\rangle_{x^{(k')}}\\
        &=\frac{k}{\langle k\rangle}\langle k'\langle\phi_{\beta}(x^{(k')}z)-1\rangle_{x^{(k')}}\rangle_{k'}
    \end{split} 
\end{equation}
In the fourth line we resummed by degree, recognising the proportion of species with degree $k'$ is given by the degree distribution. In the final two lines we relabel $k_{i}\to k$ wlog. 

Since at the fixed point for mutualistic intensity distributions $P_{\beta}$ we have $x^{(k)}=1+\eta^{(k)}$ we can make the functional equation self-consistency explicit

\begin{equation}
\begin{split}
    F_{\eta^{(k)}}(z)  &=\frac{k}{\langle k\rangle}\langle k'\langle\phi_{\beta}((1+\eta^{(k')}))z)-1\rangle_{\eta^{(k')}}\rangle_{k'}\\
    &= \frac{k}{\langle k\rangle}\langle k'\langle e^{-i\beta z+F_{\eta^{(k')}}(\beta z)}-1\rangle_{\beta}\rangle_{k'}
\end{split}
\end{equation}

Equivalently we can consider the local field $\eta$ independent of the degree satisfying $x^{(k)}=1+\frac{k}{\langle k \rangle} \eta$ as having the CGF

\begin{equation}
\begin{split}
    F_{\eta}(z)   &= \langle k'\langle e^{-i\beta z+\frac{k'}{\langle k \rangle}F_{\eta}(\beta z)}-1\rangle_{\beta}\rangle_{k'}
\end{split}
\end{equation}

In this case the exponent in the Griffiths phase is given by p when $\frac{\langle k^{1+p}\rangle}{\langle k \rangle}\beta^{p}=1$. Higher order degree correlations are involved in contrast with the ER case, since with the configuration model a single species' interactions are shaped not only by its own degree but also by how its neighbours' abundances are related to their own degrees.

\subsection{General mutualistic cumulants}\label{general mut cumul}
We can also find a similar criterion by looking at when there can be a solution to the integral equation. At the fixed point we have

\begin{equation}
    \begin{split}
        \langle \eta^{*r}\rangle_{C} = \langle \alpha^{r}\rangle_{C} \langle x^{*r}\rangle
    \end{split}
\end{equation}
If we consider the mutualistic system $x^{*}=1+\eta^{*}$, which is possible due to absence of extinctions, then this becomes 

\begin{equation}
    \begin{split}
        \langle \eta^{*r}\rangle_{C} &= \langle \alpha^{r}\rangle_{C} \langle (1+\eta^{*})^{r})\rangle\\
        &= \langle \alpha^{r}\rangle_{C} \sum_{k=0}^{r}\binom{r}{k}\langle \eta^{*k}\rangle
    \end{split}
\end{equation}

If we want to write this in terms of the cumulant generating functions this can be seen as 

\begin{equation}\label{eq: mut mid}
\begin{split}
    \frac{d^{r} F_{\eta}(z)}{d z^{r}}\bigg\rvert_{z=0} &=\frac{d^{r} F_{\alpha}(z)}{d z^{r}}\bigg\rvert_{z=0} \sum_{k=0}^{r}\binom{r}{k} \frac{d^{k} \exp{ F_{\eta}(z)}}{d z^{k}}\bigg\rvert_{z=0}\\
    &= \frac{d^{r} F_{\alpha}(z)}{d z^{r}}\bigg\rvert_{z=0}  \frac{d^{r} \exp{(z+ F_{\eta}(z))}}{d z^{r}}\bigg\rvert_{z=0}
    \end{split}
\end{equation}

Here we can see how each cumulant requires all lower orders recursively.

Comparing coefficients of $z^{k}$ in equation \ref{eq: mut mid} (essentially giving a power series ansatz for $F_{\eta}(z)$)  we get for the generic mutualistic system the recursion relation using again Bell polynomials
\begin{equation}
    \begin{split}
        \langle \eta^{r}\rangle_{C} &= \langle \alpha^{r}\rangle_{C}B_{r}(1+\langle \eta\rangle,\langle \eta^{2}\rangle_{C},\dots,\langle \eta^{r}\rangle_{C}) 
    \end{split}
\end{equation}

from which we can write explicitly

\begin{equation}
    \langle \eta\rangle = \frac{\langle \alpha\rangle}{1-\langle\alpha\rangle}
\end{equation}
and for $r>1$
\begin{equation}
    \begin{split}
        \langle \eta^{r}\rangle_{C} &= \frac{\langle \alpha^{r}\rangle_{C}}{1-\langle \alpha^{r}\rangle_{C}}B_{r}(1+\langle \eta\rangle,\langle \eta^{2}\rangle_{C},\dots,\langle \eta^{r-1}\rangle_{C},0) 
    \end{split}
\end{equation}

We note only the first argument of the Bell polynomial has a $1+$, with origin in the constant part of the growth rate. We see there is a divergence in the $r$th cumulant of the noise at the fixed point if the $r$th limiting cumulant of the interactions is 1. Not only, but all higher cumulantly will consequently also diverge due to their sequential dependence.

This expression is not simple to resum to be able to find the noise distribution or SAD except in certain cases, such as the Poisson. However we can still see how general the threshold of $\langle \alpha^{r}\rangle_{C}=1$ is, in terms of rendering the CGF non-analytic.

\subsection{Competitive background}\label{sec: competitive interactions}

If instead we consider a mean regularisation such that the interaction distribution is

\begin{equation}
    \begin{split}
        P_{S}(\alpha) = (1-\frac{c}{S})\delta(\alpha-\frac{\hat{\mu}}{S}) + &\frac{c}{S}\delta(\alpha-\beta)
    \end{split}
\end{equation}

Since the bulk $~S$ of interactions will take its value, this background mean must be scaled as $1/S$ to not trivially dominate and so we can have a well defined thermodynamic limit. Note the strong interactions are still mutualistic. If $\hat{\mu}>0$ there is a trivial shift in the phase diagram of the $c\beta=1$ line to $c\beta+\hat{\mu}=1$ with no qualitative difference. If instead $\hat{\mu}<0$, we have a negative background with rarer strong positive interactions. This can introduce some different behaviour, since extinctions become a possibility, and with them non-trivial instability.

This gives clearly a limiting interaction CGF $F_{\alpha}(z) = -i\hat{\mu} z + c (e^{-i \beta z}-1)$. Consequently this gives the functional equation for the CGF of noise for $c\beta<1$

\begin{equation}
    \begin{split}
        F_{\eta}(z) = -i \hat{\mu} z (1+\frac{d F_{\eta}}{-id z}\bigg\rvert_{z=0}) + c (e^{-i \beta z+F_{\eta}(\beta z)}-1)
    \end{split}
\end{equation}

Here the lowest abundance will be a species that does have any strong interactions. Hence it will have abundance $1+\hat{\mu} \langle x \rangle = 1+\frac{\hat{\mu}}{1-(\hat{\mu}+c\beta)} = \frac{1-c\beta}{1-(\hat{\mu}+c\beta)}$. This means extinctions will only be possible for $c\beta>1$ i.e. independent of $\hat{\mu}$. Naturally this is true strictly speaking only in the thermodynamic limit, though it seems numerically robust. Convergence of the mean seems more $S$-dependent - too small $\hat{\mu}$ will not converge fast enough numerically even if the regime should have a finite fixed point. 

The unbounded growth phase for which $\langle x\rangle$ is undefined/diverges, exists beyond the possibility of extinction. Thus it is given self-consistently by $\langle \eta\rangle=\langle\alpha\rangle\langle x\rangle_{\eta}$. The boundary of the phase diagram in the main text was found by solving for the line $\frac{1}{\langle x\rangle_{\tilde{\eta}}}=0$ in terms of parameters parameters $c,\beta$, where the average over $\tilde{\eta}$ is making the approximation $F_{\tilde{\eta}}(z)=F_{\alpha}(z\langle x \rangle_{\tilde{\eta}})$.


\end{document}